\documentclass[5p,12pt,authoryear]{elsarticle}
\usepackage{graphicx}
\usepackage{subfig}
\usepackage[dvipsnames]{xcolor}
\usepackage{hyperref}
\usepackage{float}
\usepackage{xcolor}
\usepackage[inline]{enumitem}
\usepackage{listings}
\usepackage{hyperref}
\usepackage{xspace}
\usepackage{txfonts}
\newcommand{\der}[0]{\mathrm{d}}
\newcommand{\swift}{\textsc{Swift}\xspace}
\newcommand{\csds}{\textsc{CSDS}\xspace}
\newcommand{\MPI}{\texttt{MPI}\xspace}

\DeclareFixedFont{\ttb}{T1}{txtt}{bx}{n}{12} 
\DeclareFixedFont{\ttm}{T1}{txtt}{m}{n}{12}  

\usepackage{color}
\definecolor{deepblue}{rgb}{0,0,0.5}
\definecolor{deepred}{rgb}{0.6,0,0}
\definecolor{deepgreen}{rgb}{0,0.5,0}
\newcommand{\matthieu}[1]{}

\newcommand\pythonstyle{\lstset{
language=Python,
basicstyle=\ttfamily,
commentstyle=\color{deepgreen},
keywordstyle=\ttb\color{deepblue},
emph={MyClass,__init__},          
emphstyle=\ttb\color{deepred},    
stringstyle=\color{deepgreen},
frame=tb,                         
showstringspaces=false,            %
keepspaces=true,
morekeywords={as, with},
}}

\lstnewenvironment{python}[1][]
{
\pythonstyle
\lstset{#1}
}
{}

\newcommand\ownsummary[2]{}

\newcommand\pythonexternal[2][]{{
\pythonstyle
\lstinputlisting[#1]{#2}}}

\begin{document} 
\bibpunct{(}{)}{;}{a}{}{,}

\title{Continuous Simulation Data Stream: \\ A dynamical timescale-dependent
  output scheme for simulations
}
\author{Lo\"ic Hausammann\corref{cor1}%
  \fnref{fn1}}
\ead{loic.hausammann@protonmail.com}
\address{Laboratoire d'Astrophysique, Ecole Polytechnique F\'ed\'erale de Lausanne (EPFL), 1290 Sauvergny, Switzerland}

\author{Pedro Gonnet}
\address{Google Switzerland, 8002 Z\"urich, Switzerland}
\author{Matthieu Schaller}
\address{Lorentz Institute for Theoretical Physics, Leiden University, PO Box
  9506, NL-2300 RA Leiden, The Netherlands}
\address{Leiden Observatory, Leiden University, PO Box 9513, NL-2300 RA Leiden, The Netherlands}

\cortext[cor1]{Corresponding author}


\begin{abstract}
  Exa-scale simulations are on the horizon but almost no new design for the
  output has been proposed in recent years.  In simulations using individual
  time steps, the traditional snapshots are over resolving particles/cells with
  large time steps and are under resolving the particles/cells with short time
  steps.  Therefore, they are unable to follow fast events and use efficiently
  the storage space.
  The Continuous Simulation Data Stream (\csds) is designed to decrease this
  space while providing an accurate state of the simulation at any time.
  It takes advantage of the individual time step to ensure the same relative
  accuracy for all the particles.  The outputs consist of a single file
  representing the full evolution of the simulation.  Within this file, the
  particles are written independently and at their own frequency.  Through the
  interpolation of the records, the state of the simulation can be recovered at
  any point in time.
  In this paper, we show that the \csds can reduce the storage space by 2.76x
  for the same accuracy than snapshots or increase the accuracy by 67.8x for the
  same storage space whilst retaining an acceptable reading speed for analysis.
By using interpolation between records, the \csds provides the state of the
simulation, with a high accuracy, at any time.  This should largely improve the
analysis of fast events such as supernovae and simplify the construction of
light-cone outputs.
\end{abstract}


\maketitle

\section{Introduction}

With the arrival of the era of exa-scale computing, scientists across all
domains have been focused on improving the performances of their software.  They
have been mostly looking at the scaling of their own physical computation
\citep[examples in astrophysics include
][]{schaller_swift:_2016,jetley_massively_2008,potter_pkdgrav3_2017,adams_scaling_2009,muller_escape_2019}
with, so far, little attention given to the way of writing the outputs.  Indeed,
in recent years, the HDF5 format \citep{hdf5_library} has become the de-facto
standard with some applications also relying on custom-made binary files
\citep[e.g.][]{maksimova_2021,potter_pkdgrav3_2017}. Scientists rely often
solely on the improvements within the HDF5 library itself or on the hardware for
their own code.  This approach is generally successful with simulations writing
data in parallel from 1000s of inter-connected nodes.  This format is
particularly well adapted for snapshots that simply write the state of the
simulation at a few discrete times
\citep[e.g.][]{nelson_illustris_2015,norman_simulating_2007} containing up to
trillions of particles
\cite{potter_pkdgrav3_2017,maksimova_2021,2016MNRAS.456.2361B}.  Whilst some
studies or software developments have been focused on improving the performances
of the I/O
\citep[e.g.][]{xiao_co-located_2012,ross_visualization_2008,ma_high-level_2006,mitra_efficient_2005,2017A&C....20...52R},
or on increasing the level of abstraction
\citep[e.g.][]{godoy_adios_2020,luttgau_toward_2018,zheng_flexio_2013,abbasi_extending_2009},
little work has been done on rethinking the general framework used to store data
from simulations.

An interesting case where such rethinking was done is the construction of
light-cone data.  Such outputs reproduce the behavior of observations, where
looking further away corresponds to looking back in time.  As only the particles
contained within the light cone surface are interesting, techniques have been
developed to increase the performance of the simulation
\citep[e.g.][]{garaldi_dynamic_2020} and to write the outputs in a more
efficient way \citep[e.g.][]{evrard_galaxy_2002}.  This last technique consists
in writing a type of snapshot where the particles are written only when they
cross the light cone surface.  While this approach needed to rethink the
outputs, it is not a general solution for astrophysics.

\begin{figure}
  \begin{center}
    \includegraphics[width=0.5\textwidth]{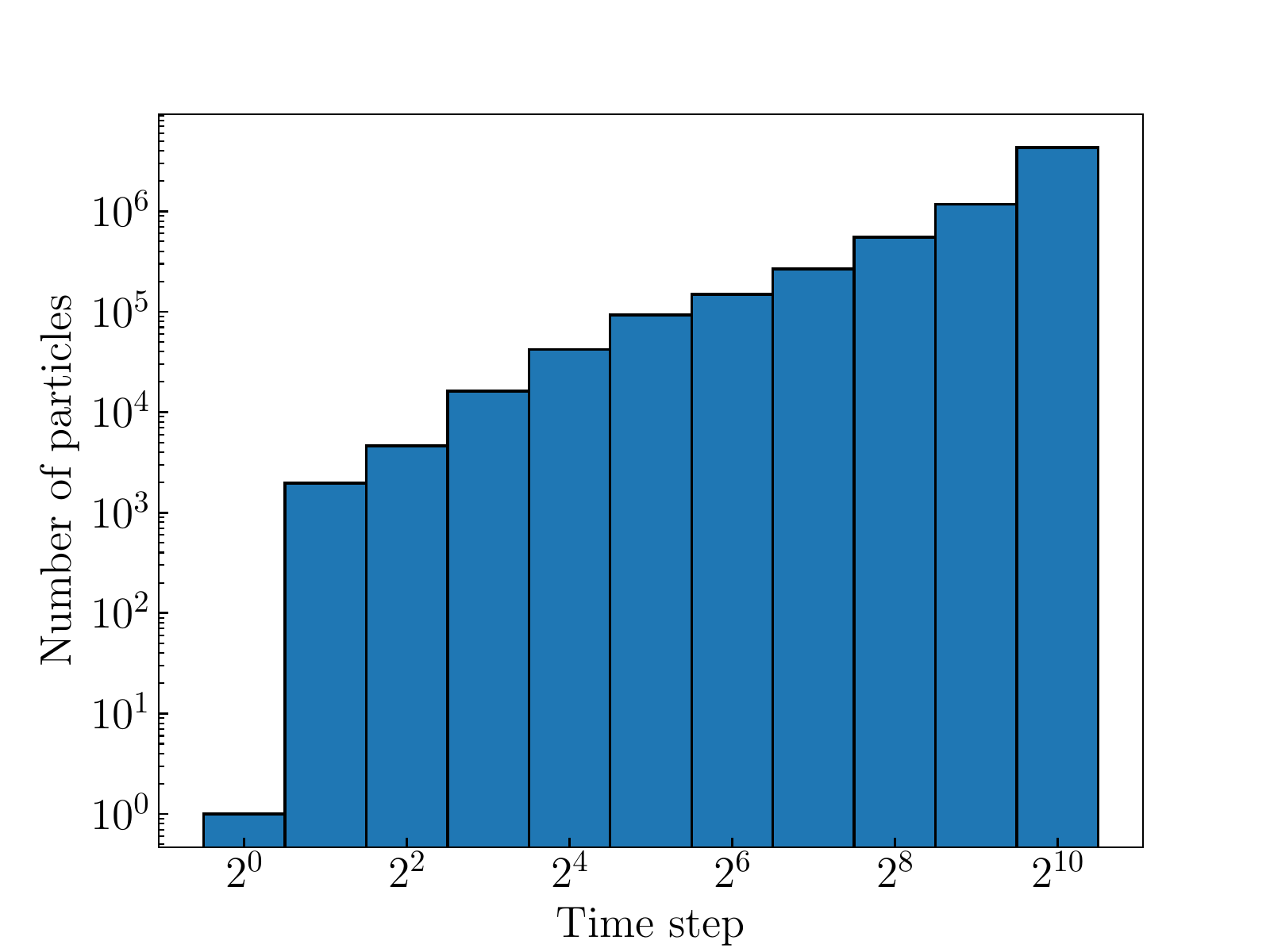}
    \caption{ Distribution of time-step sizes (normalized to the smallest
      time-step) extracted from a cosmological simulation run with the EAGLE
      model within a periodic box of 12 Mpc at $z=3$.  In this simulation, the
      ratio between the largest and smallest time steps is already above 1000
      and only a small fraction of the particles really need a high time
      resolution.  }
    \label{fig:timestep_distribution}
  \end{center}
\end{figure}

In cosmological simulations (but also more generally in other simulations where
many time-scales are coupled), the gravity is producing large difference of time
scales through the simulated volume.  As it would require too much computational
time to use a single, global, time-step size, a multi (local) and individual
time step approach was designed where each particle or cell is evolved at its
own time scale.  This approach ensures that all the particles or cells are
evolved at the same relative accuracy
\citep{aarseth_dynamical_1963,springel_cosmological_2005}.  As an example of
this, in Fig. \ref{fig:timestep_distribution}, we show the distribution of time
steps within a cosmological simulation run with the EAGLE model
\citep{schaye_eagle_2015}.  Only an almost negligible fraction of the particles
actually needs a small time step and the simulation analysis often focuses on
them. They are located in the most active regions of the simulation Even if this
example simulation is relatively small ($10^7$ particles), already a ratio of
1000 can be observed between the smallest and largest time-step sizes.  For the
outputs, it means that writing all the particles/cells together can potentially
provide a far too high accuracy for particles/cells within regions using long
time-steps (e.g. in voids) and far too low for particles/cells using very short
steps (e.g. within galaxies).  A consequence of this is that the traditional
snapshot model is unable to follow accurately fast events such as supernovae
feedback effects and render their interpretation more complicated.  Only by
writing an impossibly large number of snapshots could such events be properly
followed in large-scale simulations.  To remedy this issue, many codes are
relying on an additional set of files that describes such fast events
(e.g. writing all the supernovae into a file).  While this approach is
sufficient in many cases, it lacks a complete description of the event's
environment and its impact through time. It does also require an a priori
knowledge of what events will be of interest.

In this paper, we present a novel approach, the \emph{Continuous Simulation Data
Stream} (\csds), which allows us to recover the state of the simulation at any
time with higher accuracy and a lower storage cost than snapshots. To simplify
the discussion, in this paper, we will focus on particle based codes, but this
method could certainly be adapted for Adaptive Mesh Refinements (AMR) or to any
other general method. In practice, the \csds is implemented as a stand-alone
module around the open-source cosmological code
\swift\footnote{\url{www.swiftsim.com}}
\citep{schaller_swift:_2016,schaller_swift_2018}.

The paper is organized as follows.  In section \ref{sec:descr}, we describe in
details the theory behind the \csds.  In section \ref{sec:imp}, the
implementation within \swift is provided followed by section \ref{sec:read}
where the reading strategy is explained.  Our results are separated in two
sections (\ref{sec:results} and \ref{sec:examples}).  The first one presents the
efficiency of the \csds and the second one shows some examples of applications.

The implementation presented in this paper is fully open-source and included as
part of the \swift repository\footnote{For the \emph{csds-reader}, we used
version 1.5 (available at
\url{https://gitlab.cosma.dur.ac.uk/lhausammann/csds-reader}) in this
paper. More specifically, all the plots and numbers shown in this paper were
obtained using commit \texttt{d078e2dd} of \swift and commit \texttt{c073baf0}
of the reader code.}.

\section{The Continuous Simulation Data Stream}\label{sec:descr}

The Continuous Simulation Data Stream (\csds) is a new output strategy and
mechanism aimed at replacing or supplementing, at least partially, the
traditional snapshot system.  Its main advantage resides in the writing of each
individual resolution element in their own mini logbook. It hence allows to
capture their evolution according to their \emph{own time-scale} and not only at
the fixed time resolution set globally by traditional snapshots.  Thus, it
perfectly adapts to simulations using individual localized time-step sizes and
can achieve extremely high time resolution while keeping the output size
reasonable. Capturing the detailed progression of the elements using the
smallest time-step bin would require a prohibitive number of fixed-time
snapshots; whilst our mechanism allows us to track this evolution precisely.

\subsection{Overview of the different components}

To achieve this high resolution, the \csds uses a single file, called the
\emph{logfile}, that describes the whole evolution of the simulation.  Within
this file, all the information is written in the form of per-particle
\emph{records} (except for the header containing some general meta-data).  This
approach allows to write the particles individually and at their own timescale
without any need to synchronize them during the simulation.  When reading the
\csds for analysis, a synchronization is still required and done through an
interpolation.

To construct the logfile, a general header is written during the initialization
of the simulation.  Then, a time record for the initial step is written followed
by a particle record for all the particles in the volume (i.e. this corresponds
to the initial conditions).  The particles then carry a clock (see below) giving
them an individual dump frequency.  At each step of the simulation, the \csds
writes a \emph{time record} and a \emph{particle record} for the particles whose
timer is up.  At the end of the simulation, all the particles are written one
final time and a time record concludes the logfile.  Users can hence reconstruct
the entire history of a given particle by hoping from record to record through
the logfile. If a user requests details about a given particle in-between two
time-records, interpolation is used. The accuracy (faithfulness) of the whole
mechanism is entirely dictated by the frequency of the dump of each particle. We
note, for instance, that for the simulation shown on Fig
\ref{fig:timestep_distribution}, a large fraction of particles would not need
any records between their initial and final dump as their trajectories can be
well recovered just by interpolation.

To simplify the reading of the logfile, the \csds uses a set of {\it index
  files} that describe the state of the simulation at a given time.  While this
mimics the behavior of snapshots, they tend to be smaller as they only contain,
for each particle, the last known location within the logfile and IDs of the
particles.  The index files contain also the history of the particles created
and deleted during the simulation (e.g. for \MPI exchanges, star formation and
black hole interactions for cosmological simulations).  Theses files are
generated after the simulation and, from our experience, a small number
($\sim10$) of index files is sufficient to achieve good analysis performance.

Finally, we also include a {\it metadata file} that contains details on the
simulation (e.g. units, cosmological model, subgrid parameters). This file will
not be discussed further as the details are not important to the \csds
mechanism.

\begin{figure*}
  \begin{center}
    \includegraphics[width=\textwidth]{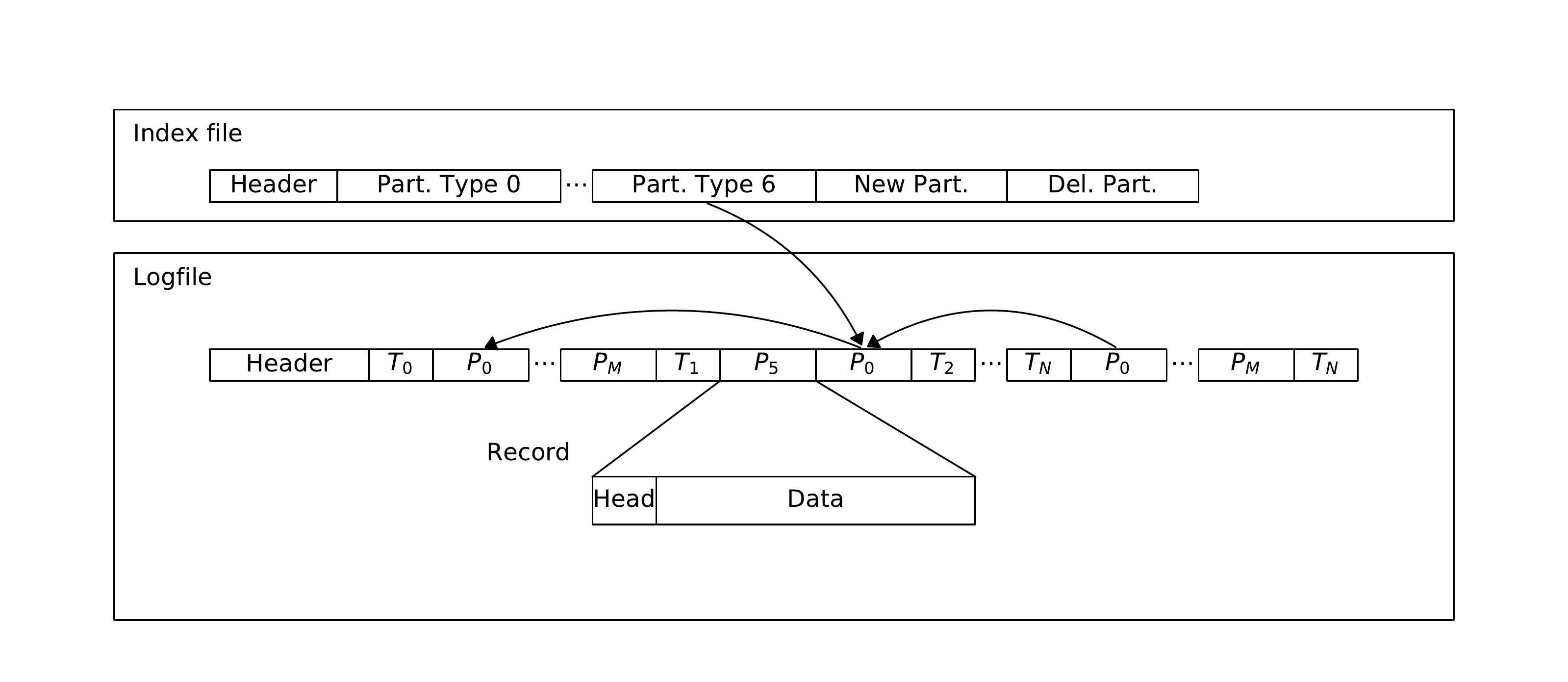}
    \caption{Representation of the different files.  The logfile contains the
      output of the simulations and the index files can be used to speed up the
      reading of the logfile.  The logfile starts with a header containing
      information about the format (e.g version numbers and the different masks,
      see Section \ref{section:record_file}).  Then the file will only contain
      records such as the one in the zoom-in.  They can either contain a time or
      a particle and are always composed of a header and the data. In the
      header, a mask is stored in order to describe the data part and also an
      offset to the previous or next corresponding record (see Section
      \ref{section:record}).  At the beginning of the simulation, we first write
      down the time ($T_0$) and the initial conditions ($P_0$ to $P_M$).  Until
      the end of the simulation, the \csds writes down the time step at the
      beginning of each step and only the required active particles (e.g. $P_0$
      and $P_5$, see Section \ref{sec:part_writing} for more information about
      the writing criterion).  Finally, the code writes the particles at the
      final time and ends up with a second writing of the time step as a
      sentinel indicating the end of file.  The index files are written once the
      simulation is done in order to speedup the reading.  They start with a
      header (e.g. simulation time, number of particles, ...) and then the
      particles sorted by type.  For each particle, the position of the last
      record (offset) and their ID is written.  At the end, the file contains
      the information about the particles created and removed since the last
      index file.  They are still written in the form of an offset and an ID.  }
    \label{fig:record_file}
  \end{center}
\end{figure*}

As a summary, the different objects used in the \csds and their relations are
shown in Fig. \ref{fig:record_file}.

\subsection{Description of the records}\label{section:record}

Each record in the logfile is made of a small header followed by the data, as
shown in the zoom-in at the bottom of Fig. \ref{fig:record_file}. In the header,
we store a \emph{mask} and an \emph{offset}.

The mask is a simple bit mask describing the fields contained in the data
(coordinates, velocities, internal energies, etc.).  This means that based only
on the bit mask, a record can be identified as containing either the current time
or a particle field. It also means that we can have different writing
frequencies for different particle fields (e.g. metallicities evolve much more
smoothly than temperatures, effectively requiring fewer entries).  The offset is
the distance in the file to the \emph{previous} record of the particle. It means
that the evolution of a particle can be quickly followed backwards thanks to
this offset.

As usually we are interested in moving forward in time, the offset of the
records are reversed during the first read of a logfile after a simulation has
completed \footnote{This operation is one of the slowest. Thus we might drop it
in the future and encourage the users to move backward in time.}.

In the data part of a record, we simply copy the fields corresponding to the
mask into the file.  As the different fields are written the one after the other
without any information in between, the order when writing and reading a record
must be respected. This order is indicated in the logfile header.

\subsection{Description of the logfile}\label{section:record_file}
\ownsummary{Aquamarine}{
  Explain the header and the timestamp strategy.
  Explain initial / final writing.
  Explain restart strategy.
  Produce an image for the logfile with a zoom into one record.
}

The logfile starts with a header that contains the version, the direction of the
offset (in order to know if they need to be reversed), the size of the strings,
the number of different masks, all the available masks (name and data size), and
finally the masks for each particle types (in the writing order).  It could be
extended with more data without risk as we also store the position of the first
record.  This general structure is shown on Fig. \ref{fig:record_file}.

When the simulation starts, the initial time is written followed by a record for
each particle presents in the initial conditions. As the \csds requires some
interpolations, this step is required in order to avoid any extrapolation.  Then
at each simulation step, the \csds writes a time record followed by as many
particle records as required by their individual write frequencies (see section
\ref{sec:part_writing} for a discussion about the writing criterion).  The
fraction of particles written during a single step can largely differ between
different steps.  On some steps, the \csds writes 0 particle due to a low number
of active particles and/or a lack of active particles requiring a log.  The
opposite case also exists where all the particles are written.  At the end of
the simulation, we write the final time followed by all the particles in order
to avoid any extrapolation and conclude with a copy of the last time record as a
sentinel marking the end of the file.

\subsection{Restart Strategy}
HPC software should expect to be killed at any time due to a system failure or to
be limited in the amount of time they can run.  In order to restart the
simulation, different strategies exist, but all relies on dumping at least part
of their memory into so-called check-point files and restarting from them.  

The \csds is no different from the rest of the software package in this situation.  While the simulation would be written to the check-point fiels as usual, the \csds simply dumps the position of the last record written in the logfile.  
When restarting, the \csds starts writing after
the position of the last record and thus automatically get rid of the
unnecessary data written between the last restart file and the crash.
We emphasize that we do not intend the \csds mechanism to be used as
a check-pointing mechanism by itself.

\subsection{Description of the index file}\label{sec:index_files}

The index files are totally optional as they contain information that exists in
the logfile. They, however, allow for an efficient usage of the logfile.  For
example, if the simulation at the final time is requested, without the index
files, it would be required to read the whole logfile in order to find all the
existing particles and then update them until reaching the final time.  To solve
this problem, the index files contain enough information to reconstruct quickly
the state of a simulation \emph{at a given time} (like a traditional snapshot
would) using the logfile.

The index files are generated when reading the logfile and are set at regular
interval in simulation time.

An index file starts with the current time step and number of particle and then
an array of ID and offset to the last record is written for each particle
(sorted by particle's type).  Next, we write down the history of the created and
suppressed particles since the last index file (if any).  The two histories are
built in the same way than the current state.  They consist first in the number
of new particles along with an array of ID and offset (sorted by type) and then
the same for the suppressed particles.

It is worth mentioning that a low number of index files requires a larger amount
of random access memory (RAM) than a large number of index files.  Indeed, the
history of all the particles created or removed in between two index files has
to be kept in memory. So increasing the number of index files can help on
machines with a low amount of memory.

\subsection{Particle writing criterion}\label{sec:part_writing}

The \csds writes down particles at different times and uses interpolations to
reconstruct a snapshot at any time in-between records. A natural consequence of
this is that the criterion for writing a particle should depends on the required
quality for the reconstruction.

A possibility is to keep in memory some information about the particle since the
last record and use this information to evaluate the quality of the future
interpolation.  For example, if one wishes to do a linear interpolation of a
field $f$ between records, the error ($\propto \mathrm{d}t^2 \max |f''(t)|$)
could be evaluated based on the time difference since last record ($dt$) and the
second derivative of $f$. This method works well with the \csds as it can be
applied independently to each particle field and then write only a subset of the
fields in each record. It would therefore reduce the size of the logfile without
losing accuracy. This is also philosophically very desirable. It means users can
specify the accuracy of the desired output rather than having to pick (almost)
arbitrary a time between traditional snapshots.

Unfortunately, this requires several new fields for each field written in the
logfile and therefore increases considerably the memory required for the
simulation.  This increase will also result in larger MPI communication and,
thus, a slow down of the simulation.

A simpler solution is to use a criterion based on the number of steps since the
last record (i.e. write every $N$ active steps). As the particle time-step size
is based on the time scale allowing to correctly integrate the particle's
properties (e.g. CFL condition), the quality of the future interpolation is
hence directly linked to the quality of the actual simulation integration.

\subsection{Parallelism considerations}

As the \csds relies heavily on the access to the logfile (both for the reading
and writing), a careful design of the I/O and good strategies to deal with
shared and distributed memory parallelism are required.  Firstly, the I/O needs
to ensure a low number of access to the data storage device in order to reduce
the impact of the storage latency.  Secondly, the \csds needs to be protected
against race conditions due to parallel processing.  The solution to these
challenges differ depending on the type of simulations (shared memory or
distributed) and are therefore described separately.

\subsubsection{Memory-mapped file}

The accesses to the logfile can dramatically decrease the performances of the
\csds due to the storage latency, therefore direct access through the system
functions \texttt{write} and \texttt{read} is strongly discouraged as they will
perform an operation on the file storage at each call.

In order to avoid this problem, operating systems (OS) provide functions to
memory-map a file and perform lazy operations on it (see e.g. \texttt{mmap} for
the POSIX standard).  This means that the OS will load new pages \footnote{A
page corresponds to a part of a file.}  only when requested. When the memory
accesses are predictable, the OS can load them in advance.  The pages will be
kept in the memory until the OS estimates that they will not be accessed
anymore.  When unloading a page, the OS lazily copies it to the file storage.

It is also worth pointing the fact that this type of memory-mapped file is very
easily manipulated through the use of a pointer. This pointer behaves exactly as
if the whole file was loaded in memory and the OS manages everything in the
background.

In case of failure or crash, all the information since the last \texttt{mmap}
synchronization will be lost, but for the rest of the file, everything will be
left intact.  In order to ensure a safe restart of the simulation, it is
possible to manually synchronize a memory-mapped file between the RAM and the
file system (\texttt{msync}).  This manual synchronization should be kept to a
strict minimum and should be called only when writing the restart files.

\subsubsection{Shared memory strategy}

In shared memory parallelization, two threads can try to access and update the
same bits at the same time which results in a race condition.  To avoid this
problem, so-called \emph{atomic} operations have been developed in most
programming languages in order to ensure that a single thread access a variable
at a given time. Unfortunately, the atomic operators slow down the execution of
the code and therefore should be used as little as possible.

For the \csds, when writing a record to the logfile, a race condition can arise
as two threads may decide to write at the same place and erase the work of the
other. To avoid this problem, the memory-mapped file does not need to be
protected, but only the pointer to the next free bytes.  When a thread needs to
write a record, it simply computes the record's size, and then increments the
pointer by the size with an atomic operator. It can later freely write on the
assigned memory. This can be done for multiple particles at a time to decrease
the amount of atomic operations.

\subsubsection{Distributed memory strategy}
Unfortunately, the simple approach highlighted above works only for shared
memory models, thus in distributed memory parallelization, a different technique
is required.

Two different possibilities are often used when dealing with files.  Either each
MPI rank owns a single file or we can predict the required amount of memory used
by each rank and can then safely write synchronously in different part of the
file.  Even if the memory could be easily predicted by checking all the active
particles before the beginning of a time step, we decided to use a single
logfile per rank for the \csds as it avoids these additional computations.

For the index files, we follow the same approach and write a set of files for
each rank.  Another advantage is that with \MPI, the simulation is usually split
in sub-volumes (domains) and each volume belongs to a single rank.  This means
that a distributed approach allows us to not read the entire logfiles when
looking at only a small volume of the simulation for analysis. All is required
is a bunch of meta-data describing which sub-volume is in which sub-file. We
note that an approach using only MPI would also be possible with one logfile 
per rank now corresponding to one logfile per compute core. The overall \csds
mechanism is agnostic of the underlying domain and parallelisation strategy.

As the different ranks are exchanging particles (e.g. when they move out of
their domain), the \csds needs to follow the particles when they are leaving /
entering a rank.  In the logfile, a particle record is written when the particle
leaves (enters) the rank and contains a flag giving the ID of the other rank.
In the index file, the particle is simply considered as being created or deleted
and thus is written in the history.

\section{Implementation in the \swift code}\label{sec:imp}

\swift \citep{schaller_swift:_2016,schaller_swift_2018} is an open source code
designed for cosmological hydrodynamical simulations but also used in planetary
science (\cite{kegerreis_planetary_2019}) and with extension to engineering
applications \citep{Bower_2022}.  We tried to implement the \csds as
independently as possible from the rest of \swift and recommend interested
readers to copy/adapt the writer from \swift into their own code.  While the
writer is obviously somewhat implementation-dependent due to the differences in
physics and implementation, the reader, however, should be quite universal.

\subsection{Specifics of the \swift code}

\swift uses MPI in-between ranks and a task-based parallelism approach relying
on pthreads \citep{gonnet_quicksched_2016} within rank. This latter approach
makes it an excellent candidate for the \csds as it is able to deal with the
\csds output whilst doing some other computations.  Thus it reduces the stress
inflicted to the storage device due to the lower number of threads accessing the
files concurrently and should provide a higher efficiency than if all the
threads were writing at the same time (as would be the case with a more
traditional parallelism approach).

\swift is designed as an HPC code that interacts the different types of
particles\footnote{In cosmological simulations, we use up to 7 different types
of particles representing the gas, dark matter (2 types), stars, black holes,
sink particles and neutrinos.} together with a minimal knowledge of the
underlying interaction models. It means that the code can easily switch between
different type of simulations (e.g. cosmology, planets or engineering). During
the development of the \csds, we also tried to follow this modular approach and
included different writing strategy for the different particles and physics
modules.

\subsection{Implementation of the \csds in \swift}

In the first sections, an overview of the \csds was given. As it does not
include some deep technical details, we will cover the most important of them in
this section.  Let us start with our management of the file size followed by
masking details.

As we cannot predict the final file size before running the simulation, we need
a mechanism to increase the file size if the expected size is not sufficient.
This is done in between every time-step where we ensure that the file is large
enough to write at least all the particles once with all their fields (even if
it does not happen).  If it is not the case, the size of the logfile is
increased by this amount times a factor in order to avoid increasing it at every
step.  This operation can be expensive for large file sizes, therefore the user
should try to estimate accurately the size and over-allocate a bit. Future
versions will attempt to make this guessing work less necessary by providing a
heuristic.  It is worth mentioning that at the end of the simulation, the file
is truncated in order to match the exact space used by the logfile.

Due to our modular approach, the masks need to be assigned in a dynamic way as
each module (gravity, SPH, etc.) defines its own fields to write. Thus,
depending on the configuration options, the number of fields can vary between
runs.  The mask are written in the first 2 bytes of the records' header; the
offsets use an additional 6 bytes.  This implies we can only have $2\times8 =
16$ masks including the ones for the time and the special flag. Therefore we
cannot log too many fields individually and have to group them into a single
mask \footnote{Depending on the future usage of the \csds, we might need to
change the masks from a field point of view to a particle type point of view
(e.g. one mask per particle type and not per field).  }.  In an effort to reduce
the number of masks needed, we check if two types of particles define the same
mask (e.g. positions) and assign the same value to both of them if it is the
case.


\subsection{Special cases}

During a simulation, the particles can go through different processes such as
type transformation, creation, or suppression. For example, in cosmological
simulations, the star formation is usually done through the transformation of a
gas particle into a star particle or could be done, in a less conventional way,
by directly creating a new star; the black holes destroy some particles through
black hole mergers or gas absorption. For engineering usages, aerodynamics
studies use wind tunnel simulations where some particles are created on one side
and removed on the opposite side in order to simulate a wind.  Finally, in the
case of non-periodic boundaries condition (e.g. planet simulations), the
particles can leave the box and therefore are removed from the simulation.

The three different cases (creation, suppression and transformation) can be
dealt with an additional mask for the particles and an additional writing of the
particles before and after each event.  In the data, we store a single integer
that includes both a flag for the type of event (creation, suppression,
transformation) and the related data (e.g. the new type of the particle).  For
the index files, we also keep in memory the suppressed and created particle
events (offset in the logfile and id of the particle) and write them in the next
index file.

\subsection{\MPI strategy}

In \swift, the volume is split in smaller volumes that are distributed to the
different ranks according to some evaluation of the work required for each
sub-volume. Throughout a simulation, the volumes tend to stay attributed to the
same MPI rank in order to avoid too much communications, therefore a particle
leaving (entering) a given volume can be considered as being removed from
(created in) a given rank without having a large impact on the logfile size. The
only difference with the suppression and deletion presented in the previous
section (i.e. due to physics events) is that we use a different flag and store
the id of the other rank involved in the transaction. We also ensure that the
offsets are correctly written between the two files.

\section{Reading strategies}\label{sec:read}

The format of the logfile and index files is not trivial and cannot be easily
read. Therefore a library is required to read the \csds and makes it more
accessible to the users.  This library (or reader) is implemented in
\texttt{C++} alongside a \texttt{python} wrapper.  A documentation is provided
in the repository.  We also provide a few functions in C in order to write the
files.  As was the case for the writer, the core implementation of the reader is
fully independent of the physics model.  Only the reading and interpolation
functions need to be updated when adding new fields.

The reader naturally provides two important possibilities: (1) follow a particle
(or a set of particles) through time in an accurate and efficient way, and (2)
generate a traditional snapshot-like output from the \csds. While the generation
of snapshots is not the best usage of the \csds, it can greatly improve the
compatibility of the \csds with existing analysis codes and provides a speedup
when the user needs to do a deep analyze of a single specific time (e.g. at
redshift $z=0$).

In order to speedup the reading process, the reader does not read directly the
logfile (except when reversing the offset of the records during the first
reading), it always starts by finding the particles in the index files and then
read the logfile starting from the last offset of the particles of interest.

When reading a subset of the particles, it is important to have an efficient way
of finding the offset of the particles. This is done by sorting the index files
according to the IDs of the particles and then using a bisection search. As the
initial order of the particles does not represent anything meaningful, the index
files can be saved once sorted.

In the case of simulations done with multiple \MPI ranks, the reader will
independently read each logfile and the corresponding index files.  It means
that the user needs to ensure that all the logfiles are read and, then,
concatenates the output together if needed. A more thorough parallelisation of
this process is left for future work.

\subsection{Implementation details}
In this section, the different steps done by the reader are summarized in order
to give an overview of the method. We focused on the most important pieces.

The reader starts by reading the header of the logfile and checks if the offsets
are already in the correct direction.  If it is not yet the case, the reversal
operation is done immediately.  This is done by starting with the first record
in the logfile.  If a previous record exists, its offset is modified in order to
point to the current record.  Once that record has been processed, the reader
moves to the end of the current record and starts over with the next record.
This requires a single linear reading of the file with random accesses to some
previous part of the file.  It is worth mentioning that, in its current
implementation, if this process is interrupted, the file will be corrupted and
it will be complicated to restore it.

Once the offsets have been reversed, the reader reads all the time records and
populate a lookup table with them in order to have a quick access to the time of
the particle records.  As this operation can be long depending on the file
size\footnote{Comparable time to reading the entire state of the simulation
(3.0s for the time array and 4.6s for the state in the millennium simulation with
$N=384^3,\, \Delta n = 10$ shown in section \ref{sec:results})}, the array is
saved at the end of the first index file and can be restored in future readings.
The reader finalizes its initialization by reading all the index files headers
and storing their time. This concludes the first read and post-processing of the
log files.

When requesting some data, the reader starts by setting the current time.  This
operation is done by selecting the two correct index files: the first one gives
the information about the last known state of the simulation and the second one
provides the information about the particles created and removed. Then the
number of available particles is computed from the index files for the
allocation of the output arrays. The particles are read one after the other
assuming that they are still present at the required time.  If it is not the
case, the code simply skips the current particle and goes to the next one. As
this approach could raise some issues if the index files or the logfile are
corrupted, we perform a sanity check at the end in order to verify that we have
the correct number of particles and raise an error if it is not the case.  It
would be possible to predict which particles are still present at a given time
thanks to the index files, but this would require too much work for the
particles that are still present. Note that, as we do not expect a lot of
particles to be removed from the simulation, we decided to use the approach
described rather than the prediction.

The particles are read field by field in order to be more flexible when each
field is written at a different frequency. As the file is memory mapped, we
always read the same file pages and they should stay within the memory during
the whole reading of a single particle.  Thus, even if we read more often the
file and do a bit more operations, the overall efficiency of the code stays the
same.

Once all the previous operations are done, the output arrays are given back to
the user. As we are reading the particles in the same order than the index
files, the output arrays are sorted in the same way (e.g. by particle type and
then ids).

\section{Results}\label{sec:results}

As mentioned in the introduction, our aim is to provide a way to recover the
state of the simulation at any time with a high accuracy and a low storage cost.
To demonstrate this, we used an isolated disk galaxy ran as a dark matter only
simulation containing 360'000 dark matter particles within an Hernquist
potential\footnote{The initial conditions and parameters are fully provided
within \swift's repository and corresponds to the
\texttt{IsolatedGalaxy\_dmparticles} example.}.

\begin{figure}
  \begin{center}
    \includegraphics[width=0.5\textwidth]{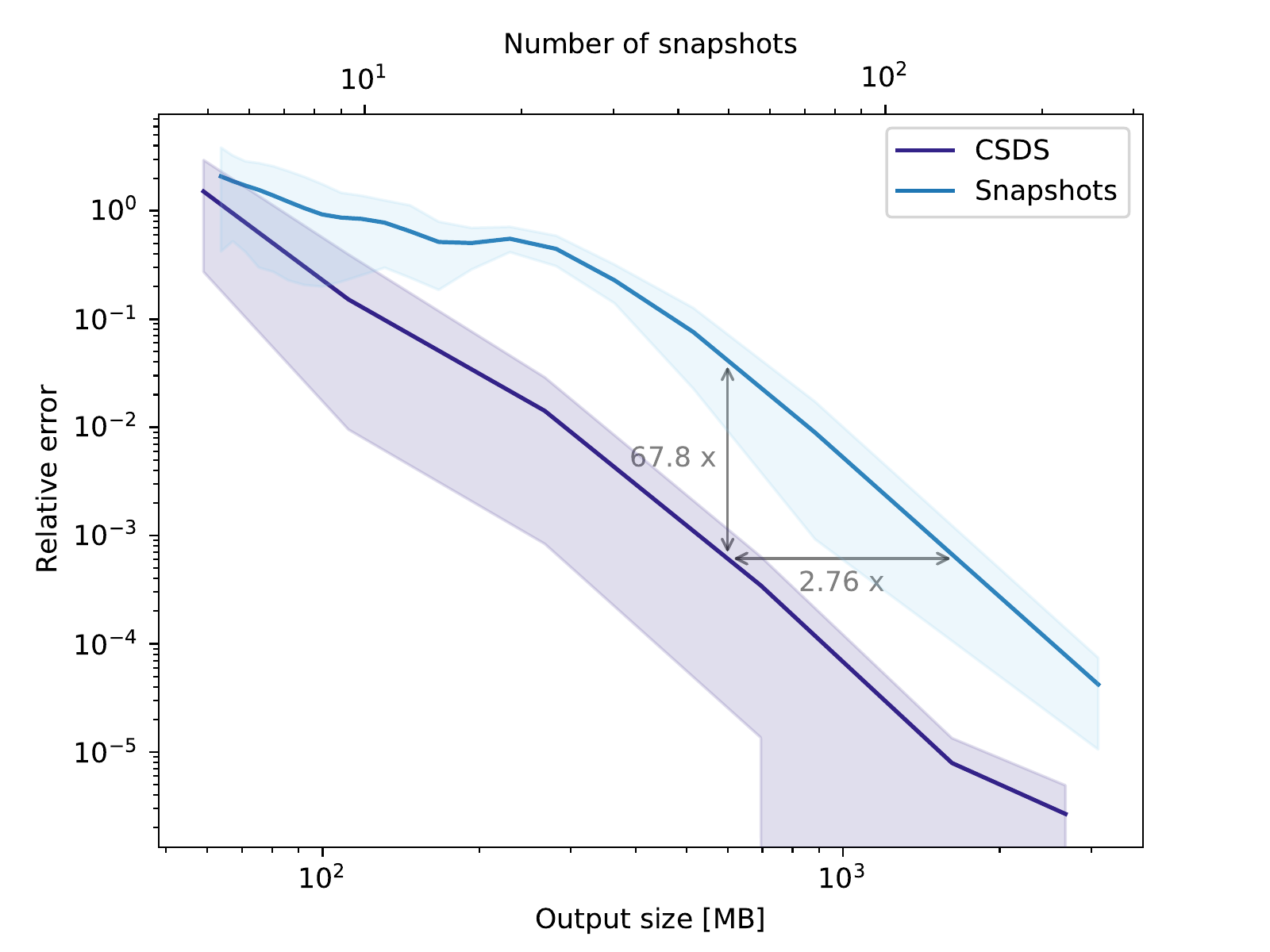}
    \caption{ Relative error of the radial position of the particles as function
      of the output size for both the traditional snapshots and the \csds.  Both
      relations are computed from the same simulation that consists of an
      isolated disk galaxy simulated only with gravity.  Only particles with $r
      \in [5, 10]$ kpc are taken into account in this figure.  The average error
      is shown as the straight lines and the distribution is shown with the
      areas representing the 16/84 percentiles.  For both the \csds and the
      snapshots, a cubic interpolation is done. The \csds requires less data
      storage for a fixed reconstruction accuracy. Or, equivalently, the \csds
      leads to a better reconstruction for the same storage.  }
    \label{fig:comparison_size}
  \end{center}
\end{figure}

Using a snapshot as our reference point, Figure \ref{fig:comparison_size} shows
the relation between the reconstruction accuracy of the particles' orbit within
the halo and the storage required for both the \csds and the traditional
snapshot strategy.  In this graph, both the snapshots and the \csds uses a cubic
Hermite interpolation based on the positions and velocities (see Appendix
\ref{sec:comoving}) to reconstruct data at intermediate times.  The accuracy is
measured by the relative error on the orbit position (here the radius).  The
relation for the snapshots is obtain through an interpolation of the reference
snapshot from two snapshots written at an equal distance in time from the
reference.  The \csds decreases the error by a factor of 67.8x at a given
storage cost or decreases the storage cost by a factor of 2.76x for a given
error.

\begin{figure}
  \begin{center}
    \includegraphics[width=0.5\textwidth]{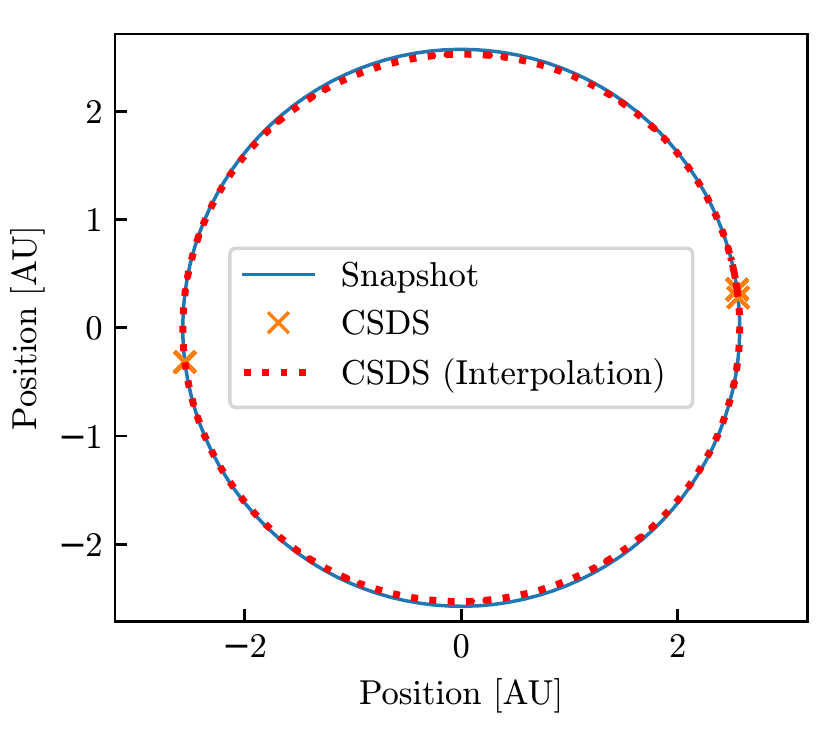}
    \caption{Orbit of a planet around a star reconstructed with the \csds (red
      line and orange crosses) and the snapshots (blue line).  The orange
      crosses correspond to the place where the \csds writes a record.  The
      quintic Hermite interpolation (using the velocity and acceleration) allows
      to accurately reconstruct the orbit with a small number of points.  }
    \label{fig:orbits}
  \end{center}
\end{figure}

Let us now look at the convergence of the \csds.  In this case, the
interpolation of the position is done through a quintic Hermite interpolation.
Due to its high convergence order, it does not require many points to accurately
represent the orbit of a particle.  In order to illustrate this point with the
\csds, on Fig. \ref{fig:orbits} we show a simulation of a planet in orbit around
a star.  The interpolation is done with the help of the velocity and
acceleration to constrain the two first derivatives in the interpolation.  While
the \csds is set at low time resolution, the traditional snapshots were written
at high resolution to allow for a fair comparison.  During a single orbit, only
three records (in orange) were written. As expected, the positions of the orange
crosses perfectly match the positions of the snapshots (in blue). The
interpolation given in red slightly differs with the snapshots far away from the
records, but provides overall a correct interpolation.  For cosmological
simulations, the situation is more complicated than in this example, but we can
already see here that a low number of points are sufficient to properly
reconstruct the orbits.

\begin{figure}
  \begin{center}
    \includegraphics[width=0.5\textwidth]{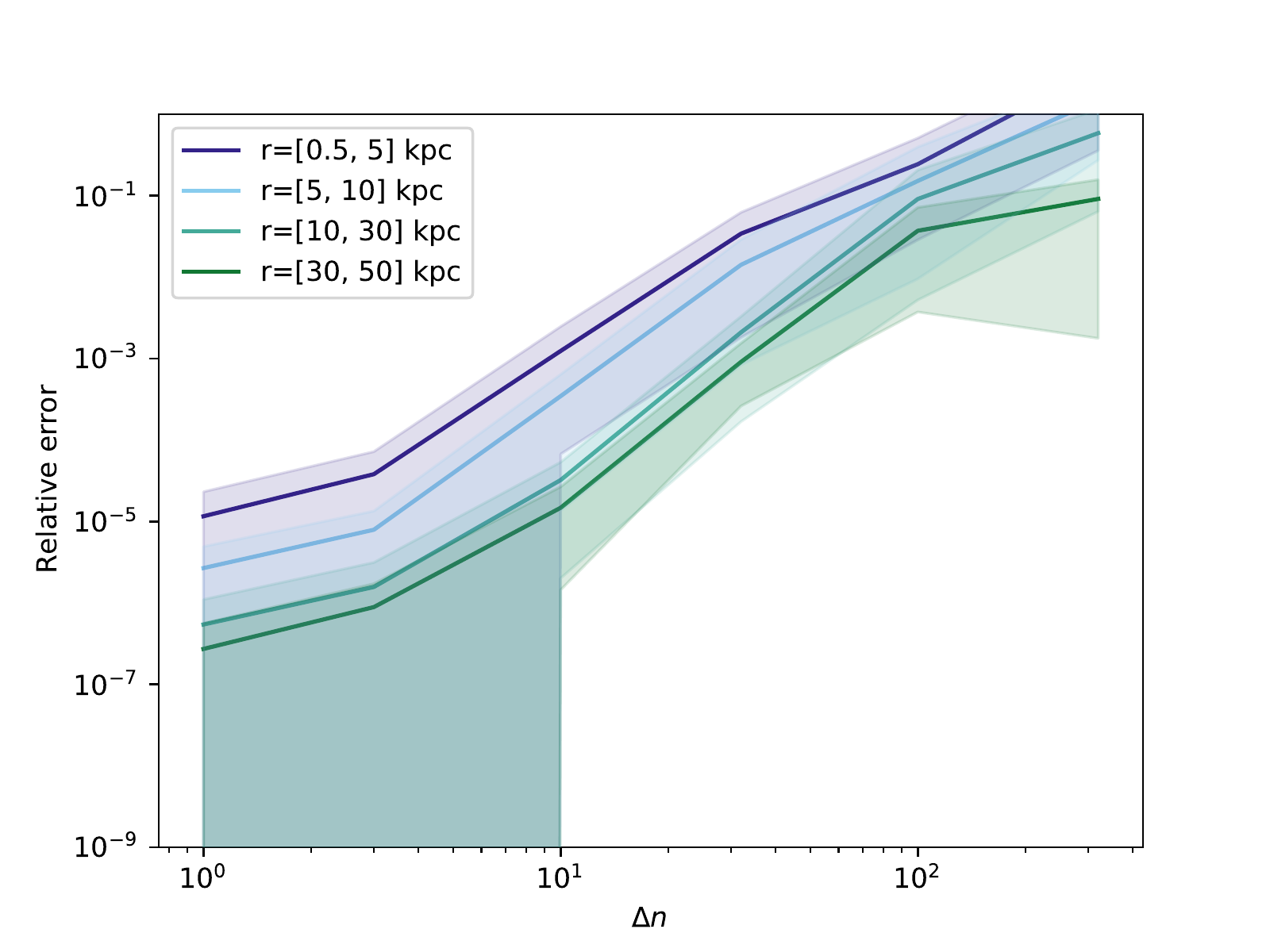}
    \caption{ Relative error of the radial positions of the particles as
      function of the number of steps between two dumping of records for an
      isolated disk galaxy ran only with gravity.  The interpolation is done
      through a quintic Hermite interpolation using the positions, velocities,
      and accelerations stored in each record.  The means of the distributions
      are shown with the solid lines and their 16/84 percentiles are indicated
      using shaded areas.  As expected, the \csds converges towards the solution
      provided by a snapshot.  }
    \label{fig:accuracy}
  \end{center}
\end{figure}

Using the isolated galaxy simulation previously described, on
Fig. \ref{fig:accuracy} we show the relative error of the \csds as function of
the number of steps between writing ($\Delta n$) for different bins of distances
from the galaxy center.  As one can expect, the \csds converges towards the
correct solution when increasing the output frequency (meaning decreasing
$\Delta n$). This relation can be used by users to set a reasonable value of
$\Delta n$ for their own simulations.

\begin{figure}
  \begin{center}
    \includegraphics[width=0.5\textwidth]{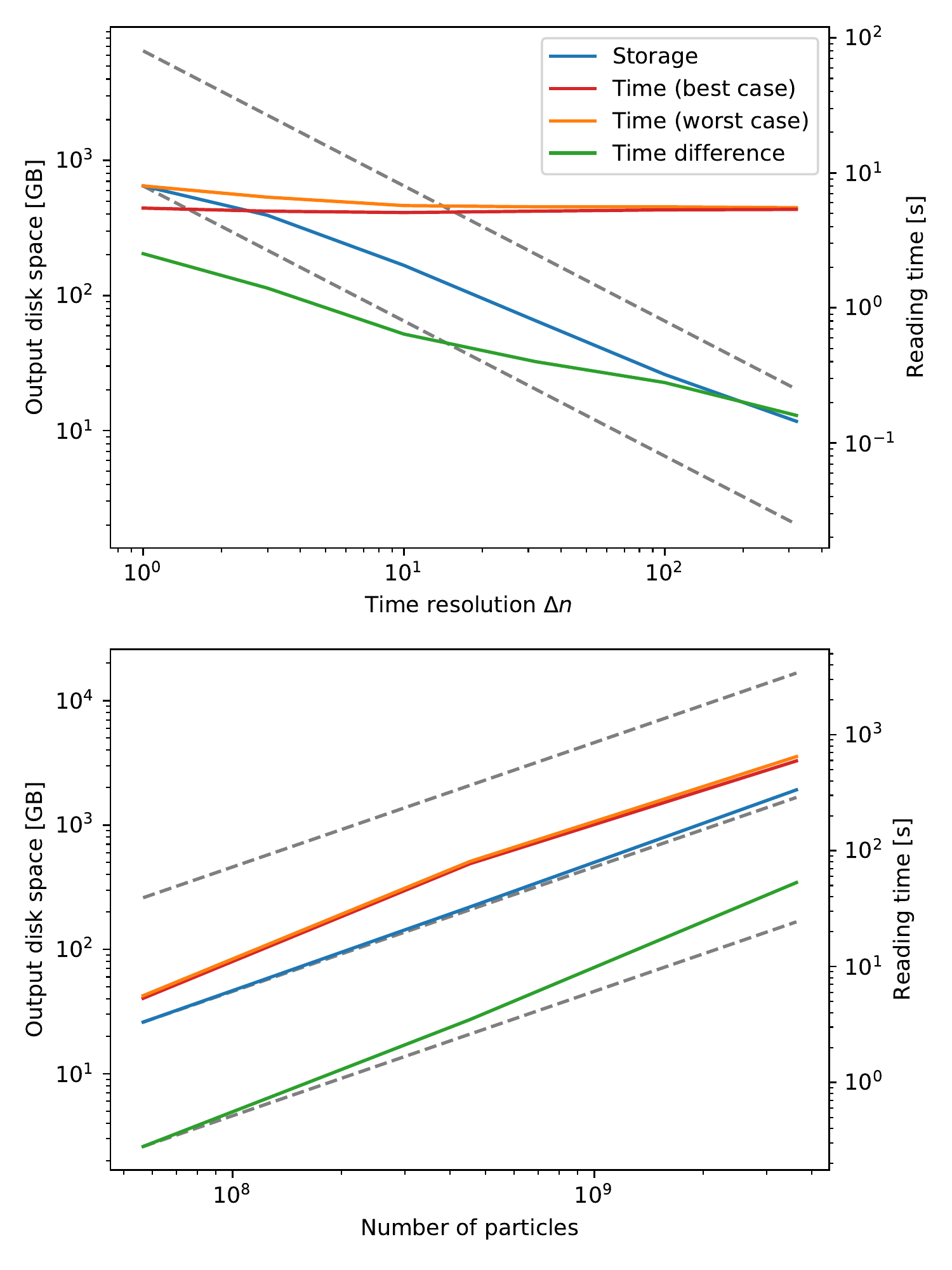}
    \caption{ Scaling of the \csds both in term of output size and reading time
      for a small version of the simulation presented by     \cite{Baugh_2019}.  
      For the first graph,
      the number of particles is $384^3$ and, for the second graph, $\Delta n =
      100$.  The first graph shows them as function of the time resolution of
      the output (number of steps between records $\Delta n$).  The second graph
      shows them as function of the simulation size.  The dashed black line
      corresponds to a linear scaling with a factor of 10 between them.  The
      best case scenario is a requested time that corresponds to an index file.
      The worst case scenario is a requested time that is just before an index
      file.  In green, the difference between the best case and the worst case
      is shown.  }
    \label{fig:scaling}
  \end{center}
\end{figure}

Finally, we demonstrate the use of the \csds on a real production simulation.
On Fig. \ref{fig:scaling} we show the scaling in term of output size and reading
speed as function of the time resolution of the \csds ($\Delta n$) and the
number of particles for a cosmological volume.  More precisely, this is the
\texttt{PMillennium-384} example provided within the \swift code.  It contains
between $N=384^3$ and $N=1536^3$ particles within a volume of $800^3~{\rm
  Mpc}^3$; i.e. a low-resolution version of the P-Millennium simulation of
\cite{Baugh_2019}.  The number of particles within the first graph is $384^3$
and the \csds time resolution in the bottom panel is $\Delta n = 100$ steps.
The largest simulation ran is within the same volume but with $N=1536^3$
particles and with exactly the same physics.  As the reading speed is dependent
on the index files, two different lines are shown.  The best case scenario (in
red) corresponds to the case where the requested time matches an index file
($z\approx 0.12$).  The worst case scenario (in orange) corresponds to the case
where the requested time is just below an index file ($z \approx 0$).  It means
that the index file used is the same as in the best case scenario, but the
particles need to be updated over 1.6 Gyr.  For the \MPI scaling, we have seen
some relatively modest changes in term of performances. On the $N=384^3$, moving
from 1 rank to 16 ranks produced a storage increase of 5\% and a variation of
reading speed within the error margin (1\%).

We note that the simulations here are relatively small ($\lesssim 10^9$ particles)
when compared to state-of-the-art simulations \citep[e.g.][]{2016MNRAS.456.2361B,Uchuu}, 
however, the data ratios between the \csds and snapshots approaches remains constant
when increasing the simulation volume. We hence expect the approach to carry over
and increase linearly with the number of particles. 

\section{Discussion}
While the snapshots and the \csds are writing the same information, the format
is far too different (single time vs whole simulation) to have a fair and
complete comparison. To make it even worse, they will not behave in the same way
depending on the type of simulations (e.g. only hydrodynamics vs with gravity)
due to the differences in the hierarchy of time-step sizes and variability of
the fields to write.  The gravity produces strong differences in timescales and
thus will improve the \csds performances when compared to hydrodynamics
simulations without any gravity that can be easily described by a single time
step.  It means that the \csds performances will strongly depend on the type of
simulation.  As the \csds was developed for simulations including gravity, we
mainly focused on such simulations in this paper.

In Figure \ref{fig:comparison_size}, we tried to show the relation between the
accuracy and the storage cost for both the \csds and the snapshots on a
simulation that includes gravity but no hydrodynamics.  As mentioned in the
previous section, the error of the snapshots is obtained through an
interpolation.  We picked a reference snapshot at equal distance from the
snapshots used for the interpolation.  It means that the snapshot error
represents the \emph{worst} case scenario.  With the \csds, there is no global
worst case scenario, only individual ones that are represented with the upper
limit of the distribution.  The \csds clearly outperforms the snapshots in term
of accuracy (67.8x) but only moderately decreases the output storage which was
our initial aim. We note that more advanced interpolation techniques, for
instance for particles in haloes \citep[e.g.][]{Smith2022} could also be used
to improve the accuracy, or reduce the number of entries needed to reach the
same precision.

In Figure \ref{fig:accuracy}, we showed a convergence study of the relative
error with respect to the number of steps between writing.  As it can be seen,
at any point on the orbit, the \csds converges towards the correct solution.
This graph shows that the \csds is able to recover the orbits of all the
particles including the most chaotic ones close to the center of a halo.  We can
see that writing every 10 (gravitational) time-steps is sufficient to produce
positions with an accuracy of 0.1\% even close to the galaxy's center.

Finally, we looked at the scaling of the output size and reading speed as
function of both the output frequency ($\Delta n$) and the number of particles
in the simulation.  As expected, they both depend linearly on the number of
particles.  In the case of the storage cost, the scaling is also more or less
linear as function of the time resolution, but that will depend on the exact
time-step hierarchy distribution.

Let us now focus on the reading time as function of $\Delta n$.  The best case
scenario is when a user is reading an index file and then directly reading a
value from the logfile without interpolation (i.e. the data requested
corresponds directly to one of the time records). It means that the time
resolution will have no impact on it and thus the reading time is independent of
$\Delta n$.  The worst case scenario is more complex as it depends on when the
index files are written and on the hierarchy of time-step sizes.  In
cosmological simulations, most of the particles will have large time steps (see
Fig. \ref{fig:timestep_distribution}).  Therefore, the average number of
``jumps'' in the logfile required to update the particles will be relatively low
(between 1.0 and 7.2 for our different time resolutions in this example).  The
extra time required to evolve the particles (i.e. finding the correct records
and interpolating) is almost negligible in all our simulations shown here even
if the evolution interval is relatively large (1.6 Gyr).  This demonstrates the
high efficiency of the evolution of the records in comparison to a single
reading of the simulation.

In term of MPI scaling, the increase of storage is expected as all the particle
exchanges will be written into the logfile.  Thus the results will strongly
depends on the MPI splitting strategy.  In this case, at the beginning, we
distribute the particles in order to have the same number of particles on each
rank.  Then the corresponding volume is assigned to a rank and it keeps it until
the end of the simulation.  The almost constant reading time can be explained by
the fact that the storage spreads over multiple files which increase the chance
of cache hit and thus reduce the time lost in file accesses.

When comparing with traditional snapshots (e.g. 13.4s for a simulation with
$768^3$ particles), the reading time of the \csds is larger but still
manageable.  While this can be an issue, it is worth to explicitly mention that
the main advantages of the \csds are the accurate reconstruction at any time
\emph{and} the possibility to evolve the particles quickly once obtained (see
the difference between the best and worst cases).  In any case, the \csds
mechanism is still at an early development state and will certainly benefit from
optimizations in the coming years.  In the meantime, if high performance is
required (or if analysis tools require data laid out in traditional way), it is
always possible to construct a snapshot from the \csds reader and then work with
it.

Some possible optimizations for future versions include tools to help with the
analysis of a completed simulation. It is often useful to be able to select the
particles in a sub-region (e.g. in a given halo). In a snapshot-based strategy
this is often achieved by sorting the particles at the end of a simulation based
on some coarse-grained grid or space-filling curve. The same approach could
be applied to the index file in the \csds case. This would then let users rapidly
retrieve particles in regions of interest. Similarly, tools developed for public 
release of data via web portals \citep[e.g.][]{2017A&C....20...52R} could also be 
modified to retrieve the particles of interest at any point in time based on this 
index file sorting and the interpolation mechanism of the \csds.

\section{Examples of application}\label{sec:examples}

In this section we showcase some application examples that directly benefit from
the \csds output strategy. \\

A problem in the comparison of large scale simulations with observations comes
from light travel times. As in observations, the further the observed objects
are, the older we observe them to be. We need to produce the same behavior in
our simulations.  This is typically done with the method called light-cones
\citep[e.g.][]{evrard_galaxy_2002,garaldi_dynamic_2020} and the main idea behind
it is to extract slices of volume in each snapshot and to stack them together
before projecting along time.  Due to the snapshots, this can be only done
approximately or with an extremely large number of snapshots.
\cite{garaldi_dynamic_2020}solved this issue by implementing the light cone
computation directly into the simulation code and redesigned it in order to
reduce the resolution in the area outside the light-cone. Thus, a single
light-cone is selected at the beginning of the simulation and cannot be changed.
With the \csds, it is possible to generate them after the end of the simulation
and for as many light cones as desired thanks to the high time resolution in the
\csds along with the interpolation. Instead of constructing snapshots as a slice
in time, we can construct slices in space-time alongside a constant light
geodesic.  For example in Figure \ref{fig:lightcone}, a light cone image is
produced from a normal \csds output, where the position of the observer and
light-cone properties were chosen \emph{after} the completion of the run. On the
figures, the axis were switched (we project both the time and position along the
x axis) in order to show the time evolution.

\begin{figure}
  \begin{center}
    \includegraphics[width=\textheight,angle=-90]{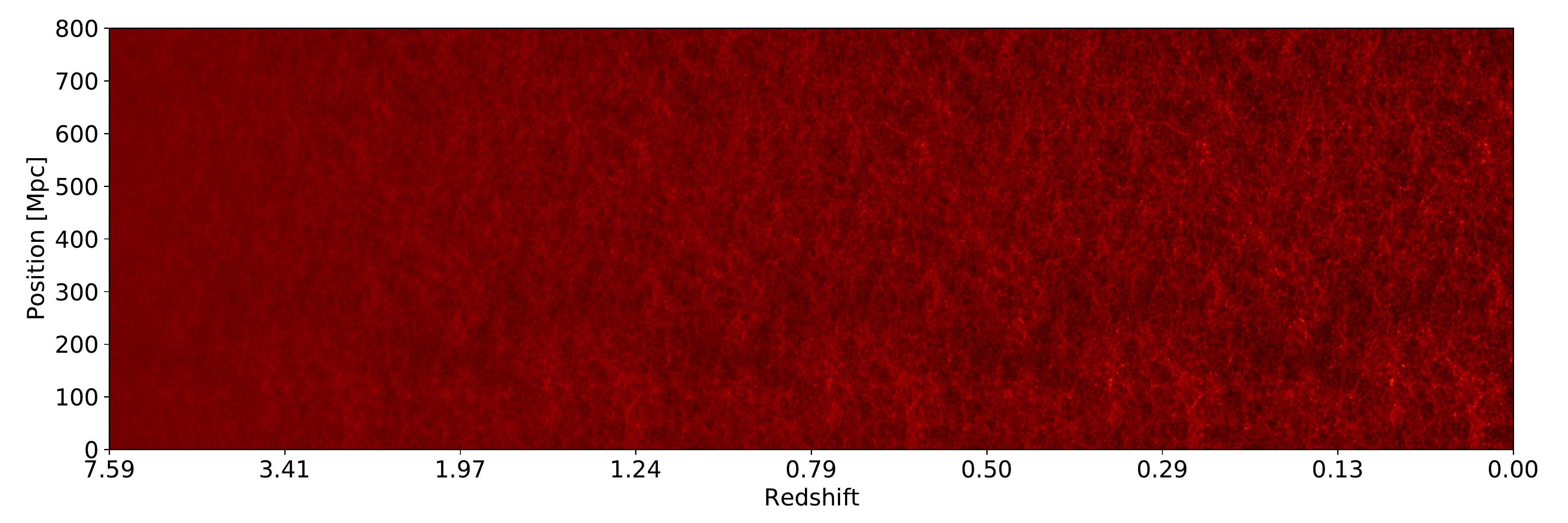}
  \end{center}
\end{figure}
\begin{figure}
    \caption{ Light cone image with the time axis along x.  The image consists
      in a single volume where the time of each particle is selected in order to
      emit lights such as it reaches the right side at z=0.  With the snapshots,
      it is required to read the particles by slices and thus produce
      discontinuities in the image.  Thanks to the \csds, this is no longer the
      case as we can find with high accuracy the time when a particle is
      crossing the light cone.}
    \label{fig:lightcone}
\end{figure}

\begin{figure}
  \begin{center}
    \includegraphics[width=0.5\textwidth]{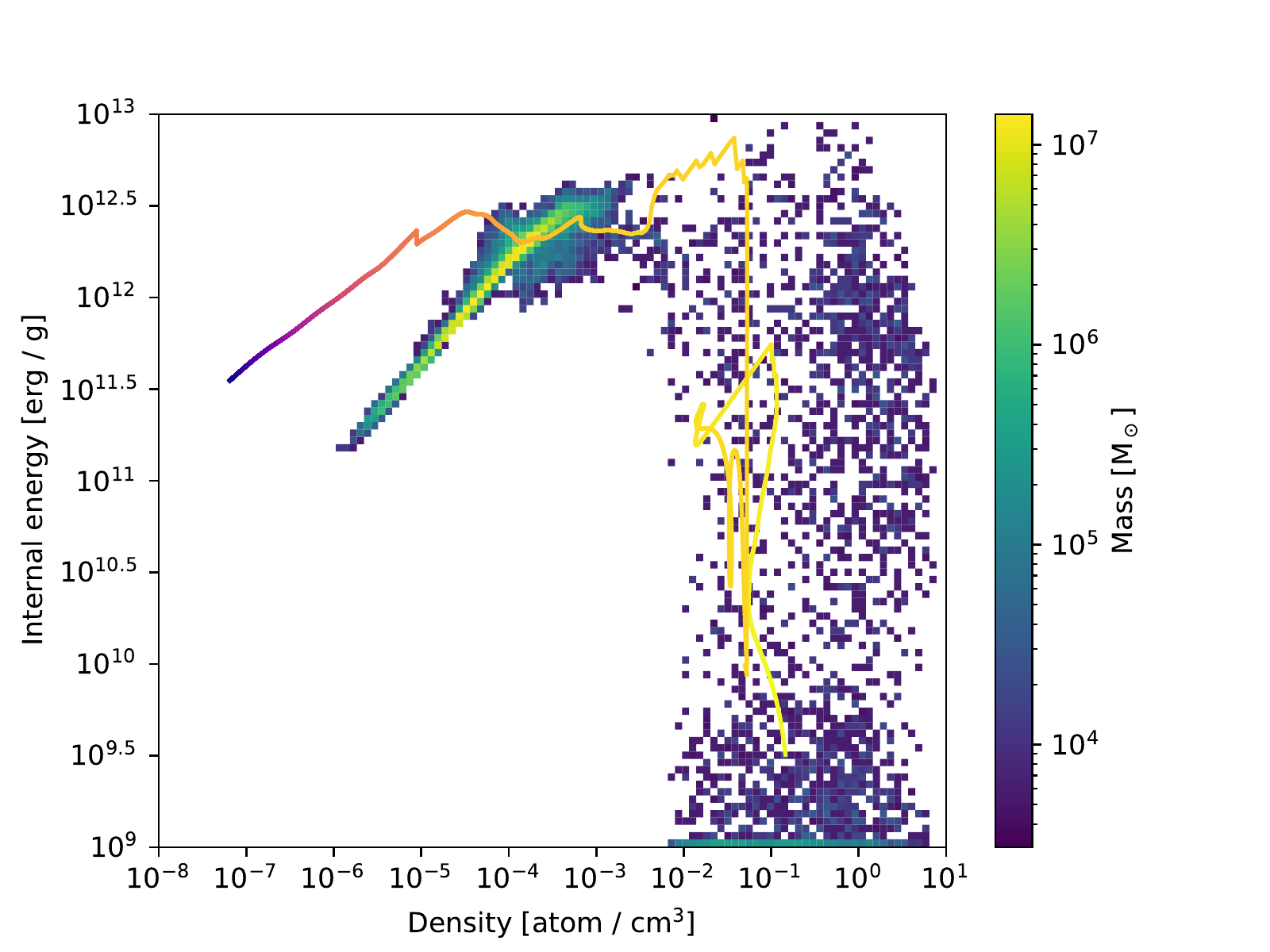}
    \caption{Phase diagram of the density and internal energy in physical units
      for a simulation of the dwarf galaxy 159 with GEAR's model
      \citep{revaz_pushing_2018} in \swift at redshift 6 for the background. The
      line shows the evolution of a particles between redshift 23 (yellow) to 4
      (purple).  With the \csds, we can accurately follow the evolution of the
      particle in the chaotic and dense medium and clearly see when a supernovae
      directly touch the particle (straight vertical line).  Due to the
      evolution between redshift 6 and 4, the final position of the particle
      followed does not match the rest of the low density particles.  }
    \label{fig:phase_space}
  \end{center}
\end{figure}

Another usage of the \csds is the study of fast events.  To illustrate this, in
Figure \ref{fig:phase_space}, a phase diagram is shown for a cosmological
simulation of a dwarf galaxies, here using the \cite{revaz_pushing_2018} model.
The diagram consists of an histogram of the distribution of the density and
internal energy of the gas at redshift 6 and the time evolution of a reference
gas particle between $z \in [23, 4]$ (yellow to purple).  The gas inside the
galaxy is in a dense phase (above $10^{-2}$ atom / cm$^3$) and can be split into
the cold (below $10^{10}$ erg / g) and hot phase that has been recently touched
by a supernova.  Inside a galaxy, the behavior of the gas is extremely chaotic
due to the constant explosion of various supernovae.  In these simulations,
supernovae are modeled by directly injecting some energy into the surrounding
gas particles and provoke a quick vertical displacement in the phase diagram.
While the timescale of the impact of a single supernovae is typically less than
a few Myr, the cosmological simulations are usually done over 14 Gyr.  The two
different timescales make it very hard to accurately follow the impact of
supernovae with the traditional snapshots.  The \csds is perfectly able to
resolve both timescales at the same time and can help to enhance our
understanding of supernovae in cosmological simulations. We note, however,
that more advanced choices for the frequency of dumps in the log might be
necessary to fully capture all the events, especially the ones around shocks
or other discontinuity in the fluid. This could, for instance, include the forceful
addition of an entry in a logfile when a shock is detected in between regularly
scheduled entries. We leave the exploration of such improvements to future work.

Finally, our last example of usage is in the production of movies.  The \csds
allows for a higher time resolution than when using traditional snapshots and is
therefore particularly useful in this case as the producer is not limited by the
available snapshots.  Thus the movie's speed can be easily adapted around
specific times or events.  Three movies have been produced for this paper and
are hosted on YouTube (described in detail on YouTube): a cosmological
simulation\footnote{The original movie without YouTube's compression is
available in the movie's description. This simulation is the same as shown in
Fig. \ref{fig:phase_space}.}
(\url{https://www.youtube.com/watch?v=OKKsk0TigNo}), the chemical evolution of
the same simulation (\url{https://www.youtube.com/watch?v=5AqmAUGndps}), and a
planetary impact (\url{https://www.youtube.com/watch?v=6aBry0CUgVw}).  While the
first and last movies show the high time resolution of the output system for a
cosmological simulation and a planetary impact, the second one shows that it can
be used directly for scientific analysis of the runs.

\section{Conclusions}\label{sec:conclusion}

We presented a new output technique that can follow the evolution of all the
resolution elements of a simulation and write them using their own individual
timescale in order to reduce the output size and increase the accuracy at any
time.  This approach is particularly well fitted for simulations that include
gravity as it produces large differences in time scales between resolution
elements. On the example presented in Figure \ref{fig:comparison_size}, we
increased the accuracy of the output by 67.8x for a given storage space (or a
factor of 2.76x in storage space for a given accuracy).  This technique largely
improves on the traditional snapshot method and will allow to improve the
analysis of fast events (e.g. supernovae) within simulations.  While this
technique has been specially created for particle-based discretization, it could
be readily adapted for grid-based techniques.

Finally, it is worth highlighting that this is a first presentation of the
method and that many improvements will be brought in the future. Among them, it
is worth to mention a few ideas that we would like to explore.  Currently, the
index files do not contain any physical information.  It could be possible to
sort the particles according to a tree in order to speedup any spatial
selection.  In most astrophysical codes, such a tree is already present in the
simulation and it would hence not require a lot of computation to produce such
tree-sorted files.  Another improvement could be done on complexifying the
frequency writing criterion.  In large scale simulations such as EAGLE, writing
the whole simulation at high resolution requires an extremely large storage.  To
reduce it, it would be possible to adapt the writing frequency according to some
physical properties (e.g. write if a given property has changed by more than
$10\%$) or spatial position (e.g. focus on a single galaxy).  This could help to
achieve an extremely high time resolution within a single galaxy.  A last
interesting improvement in term of physics could be to increase the
interpolation order by the usage of more than 2 records which would decrease
even further the storage cost for a given accuracy.


\section*{Acknowledgments}
  We thank the anonymous referees for raising interesting points which improved
  the quality of this paper.  We gratefully acknowledge the help of Bert
  Vandenbroucke and Alexei Borissov to test the \csds, as well as Jacob
  Kegerreis for help running the planetary simulation test-case.  We are also
  grateful for the ideas raised by Mladen Ivkovic, Yves Revaz and Richard Bower
  during our discussions.  We thank Alyson Brooks for reviewing an early version
  of this paper.  MS is supported by the Netherlands Organisation for Scientific
  Research (NWO) through VENI grant 639.041.749.  This work was supported by the
  Swiss Federal Institute of Technology in Lausanne (EPFL) through the use of
  the facilities of its Scientific IT and Application Support Center (SCITAS)
  and the University of Geneva through the usage of Yggdrasil.  This work used
  the DiRAC@Durham facility managed by the Institute for Computational Cosmology
  on behalf of the STFC DiRAC HPC Facility (www.dirac.ac.uk).  The equipment was
  funded by BEIS capital funding via STFC capital grants ST/K00042X/1,
  ST/P002293/1, ST/R002371/1 and ST/S002502/1, Durham University and STFC
  operations grant ST/R000832/1.  DiRAC is part of the National
  e-Infrastructure.  We are grateful to the Numpy \citep{oliphant_guide_2015},
  Matplotlib \citep{thomas_a_caswell_matplotlib/matplotlib_2018}
  SciPy \citep{jones_scipy:_2001} and IPython \citep{perez_ipython:_2007} teams
  for providing the scientific community with essential python tools.  The
  research in this paper made use of the \swift open-source simulation code
  (\url{http://www.swiftsim.com}, \cite{schaller_swift_2018}) version 0.9.0.

\bibliographystyle{elsarticle-harv}
\bibliography{csds}

\begin{thebibliography}{39}
\expandafter\ifx\csname natexlab\endcsname\relax\def\natexlab#1{#1}\fi
\providecommand{\url}[1]{\texttt{#1}}
\providecommand{\href}[2]{#2}
\providecommand{\path}[1]{#1}
\providecommand{\DOIprefix}{doi:}
\providecommand{\ArXivprefix}{arXiv:}
\providecommand{\URLprefix}{URL: }
\providecommand{\Pubmedprefix}{pmid:}
\providecommand{\doi}[1]{\href{http://dx.doi.org/#1}{\path{#1}}}
\providecommand{\Pubmed}[1]{\href{pmid:#1}{\path{#1}}}
\providecommand{\bibinfo}[2]{#2}
\ifx\xfnm\relax \def\xfnm[#1]{\unskip,\space#1}\fi
\bibitem[{Aarseth(1963)}]{aarseth_dynamical_1963}
\bibinfo{author}{Aarseth, S.J.}, \bibinfo{year}{1963}.
\newblock \bibinfo{title}{Dynamical evolution of clusters of galaxies, {I}}.
\newblock \bibinfo{journal}{Monthly Notices of the Royal Astronomical Society}
  \bibinfo{volume}{126}, \bibinfo{pages}{223}.
\newblock \URLprefix \url{http://adsabs.harvard.edu/abs/1963MNRAS.126..223A},
  \DOIprefix\doi{10.1093/mnras/126.3.223}.
\bibitem[{Abbasi et~al.(2009)Abbasi, Lofstead, Zheng, Schwan, Wolf and
  Klasky}]{abbasi_extending_2009}
\bibinfo{author}{Abbasi, H.}, \bibinfo{author}{Lofstead, J.},
  \bibinfo{author}{Zheng, F.}, \bibinfo{author}{Schwan, K.},
  \bibinfo{author}{Wolf, M.}, \bibinfo{author}{Klasky, S.},
  \bibinfo{year}{2009}.
\newblock \bibinfo{title}{Extending {I}/{O} through high performance data
  services}, in: \bibinfo{booktitle}{2009 {IEEE} {International} {Conference}
  on {Cluster} {Computing} and {Workshops}}, pp. \bibinfo{pages}{1--10}.
\newblock \DOIprefix\doi{10.1109/CLUSTR.2009.5289167}. \bibinfo{note}{iSSN:
  2168-9253}.
\bibitem[{Adams et~al.(2009)Adams, Ku, Worley, D'Azevedo, Cummings and
  Chang}]{adams_scaling_2009}
\bibinfo{author}{Adams, M.F.}, \bibinfo{author}{Ku, S.H.},
  \bibinfo{author}{Worley, P.}, \bibinfo{author}{D'Azevedo, E.},
  \bibinfo{author}{Cummings, J.C.}, \bibinfo{author}{Chang, C.S.},
  \bibinfo{year}{2009}.
\newblock \bibinfo{title}{Scaling to {150K} cores: {Recent} algorithm and
  performance engineering developments enabling {XGC1} to run at scale}.
\newblock \bibinfo{journal}{Journal of Physics: Conference Series}
  \bibinfo{volume}{180}, \bibinfo{pages}{012036}.
\newblock \URLprefix \url{http://iopscience.iop.org/1742-6596/180/1/012036},
  \DOIprefix\doi{10.1088/1742-6596/180/1/012036}.
\bibitem[{{Baugh} et~al.(2019){Baugh}, {Gonzalez-Perez}, {Lagos}, {Lacey},
  {Helly}, {Jenkins}, {Frenk}, {Benson}, {Bower} and {Cole}}]{Baugh_2019}
\bibinfo{author}{{Baugh}, C.M.}, \bibinfo{author}{{Gonzalez-Perez}, V.},
  \bibinfo{author}{{Lagos}, C.d.P.}, \bibinfo{author}{{Lacey}, C.G.},
  \bibinfo{author}{{Helly}, J.C.}, \bibinfo{author}{{Jenkins}, A.},
  \bibinfo{author}{{Frenk}, C.S.}, \bibinfo{author}{{Benson}, A.J.},
  \bibinfo{author}{{Bower}, R.G.}, \bibinfo{author}{{Cole}, S.},
  \bibinfo{year}{2019}.
\newblock \bibinfo{title}{{Galaxy formation in the Planck Millennium: the
  atomic hydrogen content of dark matter halos}}.
\newblock \bibinfo{journal}{\mnras} \bibinfo{volume}{483},
  \bibinfo{pages}{4922--4937}.
\newblock \DOIprefix\doi{10.1093/mnras/sty3427},
  \href{http://arxiv.org/abs/1808.08276}{{\tt arXiv:1808.08276}}.
\bibitem[{Bertschinger(1998)}]{bertschinger_simulations_1998}
\bibinfo{author}{Bertschinger, E.}, \bibinfo{year}{1998}.
\newblock \bibinfo{title}{Simulations of {Structure} {Formation} in the
  {Universe}}.
\newblock \bibinfo{journal}{Annual Review of Astronomy and Astrophysics}
  \bibinfo{volume}{36}, \bibinfo{pages}{599--654}.
\newblock \URLprefix \url{http://adsabs.harvard.edu/abs/1998ARA%26A..36..599B},
  \DOIprefix\doi{10.1146/annurev.astro.36.1.599}.
\bibitem[{{Bocquet} et~al.(2016){Bocquet}, {Saro}, {Dolag} and
  {Mohr}}]{2016MNRAS.456.2361B}
\bibinfo{author}{{Bocquet}, S.}, \bibinfo{author}{{Saro}, A.},
  \bibinfo{author}{{Dolag}, K.}, \bibinfo{author}{{Mohr}, J.J.},
  \bibinfo{year}{2016}.
\newblock \bibinfo{title}{{Halo mass function: baryon impact, fitting formulae,
  and implications for cluster cosmology}}.
\newblock \bibinfo{journal}{\mnras} \bibinfo{volume}{456},
  \bibinfo{pages}{2361--2373}.
\newblock \DOIprefix\doi{10.1093/mnras/stv2657},
  \href{http://arxiv.org/abs/1502.07357}{{\tt arXiv:1502.07357}}.
\bibitem[{Bower et~al.(2021)Bower, Rogers and Schaller}]{Bower_2022}
\bibinfo{author}{Bower, R.}, \bibinfo{author}{Rogers, B.D.},
  \bibinfo{author}{Schaller, M.}, \bibinfo{year}{2021}.
\newblock \bibinfo{title}{Massively parallel particle hydrodynamics at
  exa-scale}.
\newblock \bibinfo{journal}{Computing in Science Engineering} ,
  \bibinfo{pages}{1--1}\DOIprefix\doi{10.1109/MCSE.2021.3134604}.
\bibitem[{Caswell et~al.(2018)Caswell, Droettboom, Hunter, Firing, Lee,
  Stansby, de~Andrade, Nielsen, Klymak, Varoquaux, Root, Elson, Dale, May, Lee,
  Sepp{\"a}nen, Hoffmann, McDougall, Straw, Hobson, cgohlke, Yu, Ma, Vincent,
  Silvester, Moad, Katins, Kniazev, Ariza and
  W{\"u}rtz}]{thomas_a_caswell_matplotlib/matplotlib_2018}
\bibinfo{author}{Caswell, T.A.}, \bibinfo{author}{Droettboom, M.},
  \bibinfo{author}{Hunter, J.}, \bibinfo{author}{Firing, E.},
  \bibinfo{author}{Lee, A.}, \bibinfo{author}{Stansby, D.},
  \bibinfo{author}{de~Andrade, E.S.}, \bibinfo{author}{Nielsen, J.H.},
  \bibinfo{author}{Klymak, J.}, \bibinfo{author}{Varoquaux, N.},
  \bibinfo{author}{Root, B.}, \bibinfo{author}{Elson, P.},
  \bibinfo{author}{Dale, D.}, \bibinfo{author}{May, R.}, \bibinfo{author}{Lee,
  J.J.}, \bibinfo{author}{Sepp{\"a}nen, J.K.}, \bibinfo{author}{Hoffmann, T.},
  \bibinfo{author}{McDougall, D.}, \bibinfo{author}{Straw, A.},
  \bibinfo{author}{Hobson, P.}, \bibinfo{author}{cgohlke}, \bibinfo{author}{Yu,
  T.S.}, \bibinfo{author}{Ma, E.}, \bibinfo{author}{Vincent, A.F.},
  \bibinfo{author}{Silvester, S.}, \bibinfo{author}{Moad, C.},
  \bibinfo{author}{Katins, J.}, \bibinfo{author}{Kniazev, N.},
  \bibinfo{author}{Ariza, F.}, \bibinfo{author}{W{\"u}rtz, P.},
  \bibinfo{year}{2018}.
\newblock \bibinfo{title}{matplotlib/matplotlib v3.0.2}.
\newblock \URLprefix \url{https://zenodo.org/record/1482099#.XBDPecYo_0o},
  \DOIprefix\doi{10.5281/zenodo.1482099}.
\bibitem[{Evrard et~al.(2002)Evrard, MacFarland, Couchman, Colberg, Yoshida,
  White, Jenkins, Frenk, Pearce, Peacock and Thomas}]{evrard_galaxy_2002}
\bibinfo{author}{Evrard, A.E.}, \bibinfo{author}{MacFarland, T.J.},
  \bibinfo{author}{Couchman, H.M.P.}, \bibinfo{author}{Colberg, J.M.},
  \bibinfo{author}{Yoshida, N.}, \bibinfo{author}{White, S.D.M.},
  \bibinfo{author}{Jenkins, A.}, \bibinfo{author}{Frenk, C.S.},
  \bibinfo{author}{Pearce, F.R.}, \bibinfo{author}{Peacock, J.A.},
  \bibinfo{author}{Thomas, P.A.}, \bibinfo{year}{2002}.
\newblock \bibinfo{title}{Galaxy {Clusters} in {Hubble} {Volume} {Simulations}:
  {Cosmological} {Constraints} from {Sky} {Survey} {Populations}}.
\newblock \bibinfo{journal}{The Astrophysical Journal} \bibinfo{volume}{573},
  \bibinfo{pages}{7--36}.
\newblock \URLprefix \url{http://adsabs.harvard.edu/abs/2002ApJ...573....7E},
  \DOIprefix\doi{10.1086/340551}.
\bibitem[{Garaldi et~al.(2020)Garaldi, Nori and Baldi}]{garaldi_dynamic_2020}
\bibinfo{author}{Garaldi, E.}, \bibinfo{author}{Nori, M.},
  \bibinfo{author}{Baldi, M.}, \bibinfo{year}{2020}.
\newblock \bibinfo{title}{Dynamic zoom simulations: {A} fast, adaptive
  algorithm for simulating light-cones}.
\newblock \bibinfo{journal}{Monthly Notices of the Royal Astronomical Society}
  \bibinfo{volume}{499}, \bibinfo{pages}{2685--2700}.
\newblock \URLprefix \url{http://adsabs.harvard.edu/abs/2020MNRAS.499.2685G},
  \DOIprefix\doi{10.1093/mnras/staa2064}.
\bibitem[{Godoy et~al.(2020)Godoy, Podhorszki, Wang, Atkins, Eisenhauer, Gu,
  Davis, Choi, Germaschewski, Huck, Huebl, Kim, Kress, Kurc, Liu, Logan, Mehta,
  Ostrouchov, Parashar, Poeschel, Pugmire, Suchyta, Takahashi, Thompson,
  Tsutsumi, Wan, Wolf, Wu and Klasky}]{godoy_adios_2020}
\bibinfo{author}{Godoy, W.F.}, \bibinfo{author}{Podhorszki, N.},
  \bibinfo{author}{Wang, R.}, \bibinfo{author}{Atkins, C.},
  \bibinfo{author}{Eisenhauer, G.}, \bibinfo{author}{Gu, J.},
  \bibinfo{author}{Davis, P.}, \bibinfo{author}{Choi, J.},
  \bibinfo{author}{Germaschewski, K.}, \bibinfo{author}{Huck, K.},
  \bibinfo{author}{Huebl, A.}, \bibinfo{author}{Kim, M.},
  \bibinfo{author}{Kress, J.}, \bibinfo{author}{Kurc, T.},
  \bibinfo{author}{Liu, Q.}, \bibinfo{author}{Logan, J.},
  \bibinfo{author}{Mehta, K.}, \bibinfo{author}{Ostrouchov, G.},
  \bibinfo{author}{Parashar, M.}, \bibinfo{author}{Poeschel, F.},
  \bibinfo{author}{Pugmire, D.}, \bibinfo{author}{Suchyta, E.},
  \bibinfo{author}{Takahashi, K.}, \bibinfo{author}{Thompson, N.},
  \bibinfo{author}{Tsutsumi, S.}, \bibinfo{author}{Wan, L.},
  \bibinfo{author}{Wolf, M.}, \bibinfo{author}{Wu, K.},
  \bibinfo{author}{Klasky, S.}, \bibinfo{year}{2020}.
\newblock \bibinfo{title}{{ADIOS} 2: {The} {Adaptable} {Input} {Output}
  {System}. {A} framework for high-performance data management}.
\newblock \bibinfo{journal}{SoftwareX} \bibinfo{volume}{12},
  \bibinfo{pages}{100561}.
\newblock \URLprefix \url{http://adsabs.harvard.edu/abs/2020SoftX..1200561G},
  \DOIprefix\doi{10.1016/j.softx.2020.100561}.
\bibitem[{Gonnet et~al.(2016)Gonnet, Chalk and
  Schaller}]{gonnet_quicksched_2016}
\bibinfo{author}{Gonnet, P.}, \bibinfo{author}{Chalk, A.B.G.},
  \bibinfo{author}{Schaller, M.}, \bibinfo{year}{2016}.
\newblock \bibinfo{title}{{QuickSched}: {Task}-based parallelism with
  dependencies and conflicts}.
\newblock \bibinfo{journal}{arXiv:1601.05384 [cs]} \URLprefix
  \url{http://arxiv.org/abs/1601.05384}. \bibinfo{note}{arXiv: 1601.05384}.
\bibitem[{{Ishiyama} et~al.(2021){Ishiyama}, {Prada}, {Klypin}, {Sinha},
  {Metcalf}, {Jullo}, {Altieri}, {Cora}, {Croton}, {de la Torre},
  {Mill{\'a}n-Calero}, {Oogi}, {Ruedas} and {Vega-Mart{\'\i}nez}}]{Uchuu}
\bibinfo{author}{{Ishiyama}, T.}, \bibinfo{author}{{Prada}, F.},
  \bibinfo{author}{{Klypin}, A.A.}, \bibinfo{author}{{Sinha}, M.},
  \bibinfo{author}{{Metcalf}, R.B.}, \bibinfo{author}{{Jullo}, E.},
  \bibinfo{author}{{Altieri}, B.}, \bibinfo{author}{{Cora}, S.A.},
  \bibinfo{author}{{Croton}, D.}, \bibinfo{author}{{de la Torre}, S.},
  \bibinfo{author}{{Mill{\'a}n-Calero}, D.E.}, \bibinfo{author}{{Oogi}, T.},
  \bibinfo{author}{{Ruedas}, J.}, \bibinfo{author}{{Vega-Mart{\'\i}nez}, C.A.},
  \bibinfo{year}{2021}.
\newblock \bibinfo{title}{{The Uchuu simulations: Data Release 1 and dark
  matter halo concentrations}}.
\newblock \bibinfo{journal}{\mnras} \bibinfo{volume}{506},
  \bibinfo{pages}{4210--4231}.
\newblock \DOIprefix\doi{10.1093/mnras/stab1755},
  \href{http://arxiv.org/abs/2007.14720}{{\tt arXiv:2007.14720}}.
\bibitem[{Jetley et~al.(2008)Jetley, Gioachin, Mendes, Kale and
  Quinn}]{jetley_massively_2008}
\bibinfo{author}{Jetley, P.}, \bibinfo{author}{Gioachin, F.},
  \bibinfo{author}{Mendes, C.}, \bibinfo{author}{Kale, L.V.},
  \bibinfo{author}{Quinn, T.}, \bibinfo{year}{2008}.
\newblock \bibinfo{title}{Massively parallel cosmological simulations with
  {ChaNGa}}, \bibinfo{publisher}{IEEE Computer Society}. pp.
  \bibinfo{pages}{1--12}.
\newblock \URLprefix
  \url{https://www.computer.org/csdl/proceedings-article/ipdps/2008/04536319/12OmNvStcyx},
  \DOIprefix\doi{10.1109/IPDPS.2008.4536319}.
\bibitem[{Jones et~al.(2001)Jones, Oliphant, Peterson and
  {others}}]{jones_scipy:_2001}
\bibinfo{author}{Jones, E.}, \bibinfo{author}{Oliphant, T.},
  \bibinfo{author}{Peterson, P.}, \bibinfo{author}{{others}},
  \bibinfo{year}{2001}.
\newblock \bibinfo{title}{{SciPy}: {Open} source scientific tools for
  {Python}}.
\newblock \URLprefix \url{http://www.scipy.org/}.
\bibitem[{Kegerreis et~al.(2019)Kegerreis, Eke, Gonnet, Korycansky, Massey,
  Schaller and Teodoro}]{kegerreis_planetary_2019}
\bibinfo{author}{Kegerreis, J.A.}, \bibinfo{author}{Eke, V.R.},
  \bibinfo{author}{Gonnet, P.}, \bibinfo{author}{Korycansky, D.G.},
  \bibinfo{author}{Massey, R.J.}, \bibinfo{author}{Schaller, M.},
  \bibinfo{author}{Teodoro, L.F.A.}, \bibinfo{year}{2019}.
\newblock \bibinfo{title}{Planetary giant impacts: convergence of
  high-resolution simulations using efficient spherical initial conditions and
  {SWIFT}}.
\newblock \bibinfo{journal}{Monthly Notices of the Royal Astronomical Society}
  \bibinfo{volume}{487}, \bibinfo{pages}{5029--5040}.
\newblock \URLprefix \url{http://adsabs.harvard.edu/abs/2019MNRAS.487.5029K},
  \DOIprefix\doi{10.1093/mnras/stz1606}.
\bibitem[{Lind(2020)}]{lind_real-time_2020}
\bibinfo{author}{Lind, M.}, \bibinfo{year}{2020}.
\newblock \bibinfo{title}{Real-time quintic {Hermite} interpolation for robot
  trajectory execution}.
\newblock \bibinfo{journal}{PeerJ Computer Science} \bibinfo{volume}{6},
  \bibinfo{pages}{e304}.
\newblock \URLprefix \url{https://peerj.com/articles/cs-304},
  \DOIprefix\doi{10.7717/peerj-cs.304}. \bibinfo{note}{publisher: PeerJ Inc.}
\bibitem[{L{\"u}ttgau et~al.(2018)L{\"u}ttgau, Snyder, Carns, Wozniak, Kunkel
  and Ludwig}]{luttgau_toward_2018}
\bibinfo{author}{L{\"u}ttgau, J.}, \bibinfo{author}{Snyder, S.},
  \bibinfo{author}{Carns, P.}, \bibinfo{author}{Wozniak, J.M.},
  \bibinfo{author}{Kunkel, J.}, \bibinfo{author}{Ludwig, T.},
  \bibinfo{year}{2018}.
\newblock \bibinfo{title}{Toward {Understanding} {I}/{O} {Behavior} in {HPC}
  {Workflows}}, in: \bibinfo{booktitle}{2018 {IEEE}/{ACM} 3rd {International}
  {Workshop} on {Parallel} {Data} {Storage} {Data} {Intensive} {Scalable}
  {Computing} {Systems} ({PDSW}-{DISCS})}, pp. \bibinfo{pages}{64--75}.
\newblock \DOIprefix\doi{10.1109/PDSW-DISCS.2018.00012}.
\bibitem[{Ma et~al.(2006)Ma, Lee and Winslett}]{ma_high-level_2006}
\bibinfo{author}{Ma, X.}, \bibinfo{author}{Lee, J.}, \bibinfo{author}{Winslett,
  M.}, \bibinfo{year}{2006}.
\newblock \bibinfo{title}{High-level buffering for hiding periodic output cost
  in scientific simulations}.
\newblock \bibinfo{journal}{IEEE Transactions on Parallel and Distributed
  Systems} \bibinfo{volume}{17}, \bibinfo{pages}{193--204}.
\newblock \DOIprefix\doi{10.1109/TPDS.2006.36}. \bibinfo{note}{conference Name:
  IEEE Transactions on Parallel and Distributed Systems}.
\bibitem[{{Maksimova} et~al.(2021){Maksimova}, {Garrison}, {Eisenstein},
  {Hadzhiyska}, {Bose} and {Satterthwaite}}]{maksimova_2021}
\bibinfo{author}{{Maksimova}, N.A.}, \bibinfo{author}{{Garrison}, L.H.},
  \bibinfo{author}{{Eisenstein}, D.J.}, \bibinfo{author}{{Hadzhiyska}, B.},
  \bibinfo{author}{{Bose}, S.}, \bibinfo{author}{{Satterthwaite}, T.P.},
  \bibinfo{year}{2021}.
\newblock \bibinfo{title}{{ABACUSSUMMIT: a massive set of high-accuracy,
  high-resolution N-body simulations}}.
\newblock \bibinfo{journal}{\mnras} \bibinfo{volume}{508},
  \bibinfo{pages}{4017--4037}.
\newblock \DOIprefix\doi{10.1093/mnras/stab2484},
  \href{http://arxiv.org/abs/2110.11398}{{\tt arXiv:2110.11398}}.
\bibitem[{Mitra et~al.(2005)Mitra, Sinha, Winslett and
  Jiao}]{mitra_efficient_2005}
\bibinfo{author}{Mitra, S.}, \bibinfo{author}{Sinha, R.R.},
  \bibinfo{author}{Winslett, M.}, \bibinfo{author}{Jiao, X.},
  \bibinfo{year}{2005}.
\newblock \bibinfo{title}{An {Efficient}, {Nonintrusive}, {Log}-{Based} {I}/{O}
  {Mechanism} for {Scientific} {Simulations} on {Clusters}}, in:
  \bibinfo{booktitle}{2005 {IEEE} {International} {Conference} on {Cluster}
  {Computing}}, pp. \bibinfo{pages}{1--10}.
\newblock \DOIprefix\doi{10.1109/CLUSTR.2005.347041}. \bibinfo{note}{iSSN:
  2168-9253}.
\bibitem[{M{\"u}ller et~al.(2019)M{\"u}ller, Deconinck, K{\"u}hnlein, Mengaldo,
  Lange, Wedi, Bauer, Smolarkiewicz, Diamantakis, Lock, Hamrud, Saarinen,
  Mozdzynski, Thiemert, Glinton, B{\'e}nard, Voitus, Colavolpe, Marguinaud,
  Zheng, Van~Bever, Degrauwe, Smet, Termonia, Nielsen, Sass, Poulsen, Berg,
  Osuna, Fuhrer, Clement, Baldauf, Gillard, Szmelter, O'Brien, McKinstry,
  Robinson, Shukla, Lysaght, Kulczewski, Ciznicki, Piatek, Ciesielski, B{\l
  }a{\.z}ewicz, Kurowski, Procyk, Spychala, Bosak, Piotrowski, Wyszogrodzki,
  Raffin, Mazauric, Guibert, Douriez, Vigouroux, Gray, Messmer, Macfaden and
  New}]{muller_escape_2019}
\bibinfo{author}{M{\"u}ller, A.}, \bibinfo{author}{Deconinck, W.},
  \bibinfo{author}{K{\"u}hnlein, C.}, \bibinfo{author}{Mengaldo, G.},
  \bibinfo{author}{Lange, M.}, \bibinfo{author}{Wedi, N.},
  \bibinfo{author}{Bauer, P.}, \bibinfo{author}{Smolarkiewicz, P.K.},
  \bibinfo{author}{Diamantakis, M.}, \bibinfo{author}{Lock, S.J.},
  \bibinfo{author}{Hamrud, M.}, \bibinfo{author}{Saarinen, S.},
  \bibinfo{author}{Mozdzynski, G.}, \bibinfo{author}{Thiemert, D.},
  \bibinfo{author}{Glinton, M.}, \bibinfo{author}{B{\'e}nard, P.},
  \bibinfo{author}{Voitus, F.}, \bibinfo{author}{Colavolpe, C.},
  \bibinfo{author}{Marguinaud, P.}, \bibinfo{author}{Zheng, Y.},
  \bibinfo{author}{Van~Bever, J.}, \bibinfo{author}{Degrauwe, D.},
  \bibinfo{author}{Smet, G.}, \bibinfo{author}{Termonia, P.},
  \bibinfo{author}{Nielsen, K.P.}, \bibinfo{author}{Sass, B.H.},
  \bibinfo{author}{Poulsen, J.W.}, \bibinfo{author}{Berg, P.},
  \bibinfo{author}{Osuna, C.}, \bibinfo{author}{Fuhrer, O.},
  \bibinfo{author}{Clement, V.}, \bibinfo{author}{Baldauf, M.},
  \bibinfo{author}{Gillard, M.}, \bibinfo{author}{Szmelter, J.},
  \bibinfo{author}{O'Brien, E.}, \bibinfo{author}{McKinstry, A.},
  \bibinfo{author}{Robinson, O.}, \bibinfo{author}{Shukla, P.},
  \bibinfo{author}{Lysaght, M.}, \bibinfo{author}{Kulczewski, M.},
  \bibinfo{author}{Ciznicki, M.}, \bibinfo{author}{Piatek, W.},
  \bibinfo{author}{Ciesielski, S.}, \bibinfo{author}{B{\l }a{\.z}ewicz, M.},
  \bibinfo{author}{Kurowski, K.}, \bibinfo{author}{Procyk, M.},
  \bibinfo{author}{Spychala, P.}, \bibinfo{author}{Bosak, B.},
  \bibinfo{author}{Piotrowski, Z.P.}, \bibinfo{author}{Wyszogrodzki, A.},
  \bibinfo{author}{Raffin, E.}, \bibinfo{author}{Mazauric, C.},
  \bibinfo{author}{Guibert, D.}, \bibinfo{author}{Douriez, L.},
  \bibinfo{author}{Vigouroux, X.}, \bibinfo{author}{Gray, A.},
  \bibinfo{author}{Messmer, P.}, \bibinfo{author}{Macfaden, A.J.},
  \bibinfo{author}{New, N.}, \bibinfo{year}{2019}.
\newblock \bibinfo{title}{The {ESCAPE} project: {Energy}-efficient {Scalable}
  {Algorithms} for {Weather} {Prediction} at {Exascale}}.
\newblock \bibinfo{journal}{Geoscientific Model Development}
  \bibinfo{volume}{12}, \bibinfo{pages}{4425--4441}.
\newblock \URLprefix \url{http://adsabs.harvard.edu/abs/2019GMD....12.4425M},
  \DOIprefix\doi{10.5194/gmd-12-4425-2019}.
\bibitem[{Nelson et~al.(2015)Nelson, Pillepich, Genel, Vogelsberger, Springel,
  Torrey, Rodriguez-Gomez, Sijacki, Snyder, Griffen, Marinacci, Blecha, Sales,
  Xu and Hernquist}]{nelson_illustris_2015}
\bibinfo{author}{Nelson, D.}, \bibinfo{author}{Pillepich, A.},
  \bibinfo{author}{Genel, S.}, \bibinfo{author}{Vogelsberger, M.},
  \bibinfo{author}{Springel, V.}, \bibinfo{author}{Torrey, P.},
  \bibinfo{author}{Rodriguez-Gomez, V.}, \bibinfo{author}{Sijacki, D.},
  \bibinfo{author}{Snyder, G.F.}, \bibinfo{author}{Griffen, B.},
  \bibinfo{author}{Marinacci, F.}, \bibinfo{author}{Blecha, L.},
  \bibinfo{author}{Sales, L.}, \bibinfo{author}{Xu, D.},
  \bibinfo{author}{Hernquist, L.}, \bibinfo{year}{2015}.
\newblock \bibinfo{title}{The illustris simulation: {Public} data release}.
\newblock \bibinfo{journal}{Astronomy and Computing} \bibinfo{volume}{13},
  \bibinfo{pages}{12--37}.
\newblock \URLprefix
  \url{http://www.sciencedirect.com/science/article/pii/S2213133715000864},
  \DOIprefix\doi{10.1016/j.ascom.2015.09.003}.
\bibitem[{Norman et~al.(2007)Norman, Bryan, Harkness, Bordner, Reynolds, O'Shea
  and Wagner}]{norman_simulating_2007}
\bibinfo{author}{Norman, M.L.}, \bibinfo{author}{Bryan, G.L.},
  \bibinfo{author}{Harkness, R.}, \bibinfo{author}{Bordner, J.},
  \bibinfo{author}{Reynolds, D.}, \bibinfo{author}{O'Shea, B.},
  \bibinfo{author}{Wagner, R.}, \bibinfo{year}{2007}.
\newblock \bibinfo{title}{Simulating {Cosmological} {Evolution} with {Enzo}}.
\newblock \bibinfo{journal}{arXiv:0705.1556 [astro-ph]} \URLprefix
  \url{http://arxiv.org/abs/0705.1556}. \bibinfo{note}{arXiv: 0705.1556}.
\bibitem[{Oliphant(2015)}]{oliphant_guide_2015}
\bibinfo{author}{Oliphant, T.E.}, \bibinfo{year}{2015}.
\newblock \bibinfo{title}{Guide to {NumPy}}.
\newblock \bibinfo{edition}{2nd} ed., \bibinfo{publisher}{CreateSpace
  Independent Publishing Platform}, \bibinfo{address}{USA}.
\bibitem[{Peebles(1993)}]{peebles_principles_1993}
\bibinfo{author}{Peebles, P.J.E.}, \bibinfo{year}{1993}.
\newblock \bibinfo{title}{Principles of {Physical} {Cosmology}}.
\newblock \bibinfo{journal}{Principles of Physical Cosmology by P.J.E. Peebles.
  Princeton University Press, 1993. ISBN: 978-0-691-01933-8} \URLprefix
  \url{http://adsabs.harvard.edu/abs/1993ppc..book.....P}.
\bibitem[{Perez and Granger(2007)}]{perez_ipython:_2007}
\bibinfo{author}{Perez, F.}, \bibinfo{author}{Granger, B.E.},
  \bibinfo{year}{2007}.
\newblock \bibinfo{title}{{IPython}: {A} {System} for {Interactive}
  {Scientific} {Computing}}.
\newblock \bibinfo{journal}{Computing in Science Engineering}
  \bibinfo{volume}{9}, \bibinfo{pages}{21--29}.
\newblock \DOIprefix\doi{10.1109/MCSE.2007.53}.
\bibitem[{Potter et~al.(2017)Potter, Stadel and
  Teyssier}]{potter_pkdgrav3_2017}
\bibinfo{author}{Potter, D.}, \bibinfo{author}{Stadel, J.},
  \bibinfo{author}{Teyssier, R.}, \bibinfo{year}{2017}.
\newblock \bibinfo{title}{{PKDGRAV3}: beyond trillion particle cosmological
  simulations for the next era of galaxy surveys}.
\newblock \bibinfo{journal}{Computational Astrophysics and Cosmology}
  \bibinfo{volume}{4}, \bibinfo{pages}{2}.
\newblock \URLprefix \url{http://adsabs.harvard.edu/abs/2017ComAC...4....2P},
  \DOIprefix\doi{10.1186/s40668-017-0021-1}.
\bibitem[{{Ragagnin} et~al.(2017){Ragagnin}, {Dolag}, {Biffi}, {Cadolle Bel},
  {Hammer}, {Krukau}, {Petkova} and {Steinborn}}]{2017A&C....20...52R}
\bibinfo{author}{{Ragagnin}, A.}, \bibinfo{author}{{Dolag}, K.},
  \bibinfo{author}{{Biffi}, V.}, \bibinfo{author}{{Cadolle Bel}, M.},
  \bibinfo{author}{{Hammer}, N.J.}, \bibinfo{author}{{Krukau}, A.},
  \bibinfo{author}{{Petkova}, M.}, \bibinfo{author}{{Steinborn}, D.},
  \bibinfo{year}{2017}.
\newblock \bibinfo{title}{{A web portal for hydrodynamical, cosmological
  simulations}}.
\newblock \bibinfo{journal}{Astronomy and Computing} \bibinfo{volume}{20},
  \bibinfo{pages}{52--67}.
\newblock \DOIprefix\doi{10.1016/j.ascom.2017.05.001},
  \href{http://arxiv.org/abs/1612.06380}{{\tt arXiv:1612.06380}}.
\bibitem[{Revaz and Jablonka(2018)}]{revaz_pushing_2018}
\bibinfo{author}{Revaz, Y.}, \bibinfo{author}{Jablonka, P.},
  \bibinfo{year}{2018}.
\newblock \bibinfo{title}{Pushing back the limits: detailed properties of dwarf
  galaxies in a {LCDM} universe}.
\newblock \bibinfo{journal}{arXiv:1801.06222 [astro-ph]} \URLprefix
  \url{http://arxiv.org/abs/1801.06222}. \bibinfo{note}{arXiv: 1801.06222}.
\bibitem[{Ross et~al.(2008)Ross, Peterka, Shen, Hong, Ma, Yu and
  Moreland}]{ross_visualization_2008}
\bibinfo{author}{Ross, R.B.}, \bibinfo{author}{Peterka, T.},
  \bibinfo{author}{Shen, H.W.}, \bibinfo{author}{Hong, Y.},
  \bibinfo{author}{Ma, K.L.}, \bibinfo{author}{Yu, H.},
  \bibinfo{author}{Moreland, K.}, \bibinfo{year}{2008}.
\newblock \bibinfo{title}{Visualization and parallel {I}/{O} at extreme scale}.
\newblock \bibinfo{journal}{Journal of Physics: Conference Series}
  \bibinfo{volume}{125}, \bibinfo{pages}{012099}.
\newblock \URLprefix
  \url{https://iopscience.iop.org/article/10.1088/1742-6596/125/1/012099},
  \DOIprefix\doi{10.1088/1742-6596/125/1/012099}.
\bibitem[{Schaller et~al.(2016)Schaller, Gonnet, Chalk and
  Draper}]{schaller_swift:_2016}
\bibinfo{author}{Schaller, M.}, \bibinfo{author}{Gonnet, P.},
  \bibinfo{author}{Chalk, A.B.G.}, \bibinfo{author}{Draper, P.W.},
  \bibinfo{year}{2016}.
\newblock \bibinfo{title}{{SWIFT}: {Using} task-based parallelism, fully
  asynchronous communication, and graph partition-based domain decomposition
  for strong scaling on more than 100,000 cores}, in:
  \bibinfo{booktitle}{Proceedings of the Platform for Advanced Scientific
  Computing Conference}, pp. \bibinfo{pages}{1--10}.
\newblock \URLprefix \url{http://arxiv.org/abs/1606.02738},
  \DOIprefix\doi{10.1145/2929908.2929916}. \bibinfo{note}{arXiv: 1606.02738}.
\bibitem[{Schaller et~al.(2018)Schaller, Gonnet, Draper, Chalk, Bower, Willis
  and Hausammann}]{schaller_swift_2018}
\bibinfo{author}{Schaller, M.}, \bibinfo{author}{Gonnet, P.},
  \bibinfo{author}{Draper, P.W.}, \bibinfo{author}{Chalk, A.B.G.},
  \bibinfo{author}{Bower, R.G.}, \bibinfo{author}{Willis, J.},
  \bibinfo{author}{Hausammann, L.}, \bibinfo{year}{2018}.
\newblock \bibinfo{title}{{SWIFT}: {SPH} {With} {Inter}-dependent
  {Fine}-grained {Tasking}}.
\newblock \bibinfo{journal}{Astrophysics Source Code Library} ,
  \bibinfo{pages}{ascl:1805.020}\URLprefix
  \url{https://ui.adsabs.harvard.edu/abs/2018ascl.soft05020S}.
\bibitem[{Schaye et~al.(2015)Schaye, Crain, Bower, Furlong, Schaller, Theuns,
  Dalla~Vecchia, Frenk, McCarthy, Helly, Jenkins, Rosas-Guevara, White, Baes,
  Booth, Camps, Navarro, Qu, Rahmati, Sawala, Thomas and
  Trayford}]{schaye_eagle_2015}
\bibinfo{author}{Schaye, J.}, \bibinfo{author}{Crain, R.A.},
  \bibinfo{author}{Bower, R.G.}, \bibinfo{author}{Furlong, M.},
  \bibinfo{author}{Schaller, M.}, \bibinfo{author}{Theuns, T.},
  \bibinfo{author}{Dalla~Vecchia, C.}, \bibinfo{author}{Frenk, C.S.},
  \bibinfo{author}{McCarthy, I.G.}, \bibinfo{author}{Helly, J.C.},
  \bibinfo{author}{Jenkins, A.}, \bibinfo{author}{Rosas-Guevara, Y.M.},
  \bibinfo{author}{White, S.D.M.}, \bibinfo{author}{Baes, M.},
  \bibinfo{author}{Booth, C.M.}, \bibinfo{author}{Camps, P.},
  \bibinfo{author}{Navarro, J.F.}, \bibinfo{author}{Qu, Y.},
  \bibinfo{author}{Rahmati, A.}, \bibinfo{author}{Sawala, T.},
  \bibinfo{author}{Thomas, P.A.}, \bibinfo{author}{Trayford, J.},
  \bibinfo{year}{2015}.
\newblock \bibinfo{title}{The {EAGLE} project: simulating the evolution and
  assembly of galaxies and their environments}.
\newblock \bibinfo{journal}{Monthly Notices of the Royal Astronomical Society}
  \bibinfo{volume}{446}, \bibinfo{pages}{521--554}.
\newblock \URLprefix \url{http://adsabs.harvard.edu/abs/2015MNRAS.446..521S},
  \DOIprefix\doi{10.1093/mnras/stu2058}.
\bibitem[{{Smith} et~al.(2022){Smith}, {Cole}, {Grove}, {Norberg} and
  {Zarrouk}}]{Smith2022}
\bibinfo{author}{{Smith}, A.}, \bibinfo{author}{{Cole}, S.},
  \bibinfo{author}{{Grove}, C.}, \bibinfo{author}{{Norberg}, P.},
  \bibinfo{author}{{Zarrouk}, P.}, \bibinfo{year}{2022}.
\newblock \bibinfo{title}{{A lightcone catalogue from the Millennium-XXL
  simulation: improved spatial interpolation and colour distributions for the
  DESI BGS}}.
\newblock \bibinfo{journal}{arXiv e-prints} ,
  \bibinfo{pages}{arXiv:2207.04902}\href{http://arxiv.org/abs/2207.04902}{{\tt
  arXiv:2207.04902}}.
\bibitem[{Springel(2005)}]{springel_cosmological_2005}
\bibinfo{author}{Springel, V.}, \bibinfo{year}{2005}.
\newblock \bibinfo{title}{The cosmological simulation code {GADGET}-2}.
\newblock \bibinfo{journal}{Monthly Notices of the Royal Astronomical Society}
  \bibinfo{volume}{364}, \bibinfo{pages}{1105--1134}.
\newblock \URLprefix \url{http://adsabs.harvard.edu/abs/2005MNRAS.364.1105S},
  \DOIprefix\doi{10.1111/j.1365-2966.2005.09655.x}.
\bibitem[{{The HDF Group}(1997)}]{hdf5_library}
\bibinfo{author}{{The HDF Group}}, \bibinfo{year}{1997}.
\newblock \bibinfo{title}{{Hierarchical Data Format, version 5}}.
\newblock \bibinfo{note}{Https://www.hdfgroup.org/HDF5/}.
\bibitem[{Xiao et~al.(2012)Xiao, Shang and Wang}]{xiao_co-located_2012}
\bibinfo{author}{Xiao, Q.}, \bibinfo{author}{Shang, P.}, \bibinfo{author}{Wang,
  J.}, \bibinfo{year}{2012}.
\newblock \bibinfo{title}{Co-located {Compute} and {Binary} {File} {Storage} in
  {Data}-{Intensive} {Computing}}, in: \bibinfo{booktitle}{2012 {IEEE}
  {Seventh} {International} {Conference} on {Networking}, {Architecture}, and
  {Storage}}, pp. \bibinfo{pages}{199--206}.
\newblock \DOIprefix\doi{10.1109/NAS.2012.29}.
\bibitem[{Zheng et~al.(2013)Zheng, Zou, Eisenhauer, Schwan, Wolf, Dayal,
  Nguyen, Cao, Abbasi, Klasky, Podhorszki and Yu}]{zheng_flexio_2013}
\bibinfo{author}{Zheng, F.}, \bibinfo{author}{Zou, H.},
  \bibinfo{author}{Eisenhauer, G.}, \bibinfo{author}{Schwan, K.},
  \bibinfo{author}{Wolf, M.}, \bibinfo{author}{Dayal, J.},
  \bibinfo{author}{Nguyen, T.}, \bibinfo{author}{Cao, J.},
  \bibinfo{author}{Abbasi, H.}, \bibinfo{author}{Klasky, S.},
  \bibinfo{author}{Podhorszki, N.}, \bibinfo{author}{Yu, H.},
  \bibinfo{year}{2013}.
\newblock \bibinfo{title}{{FlexIO}: {I}/{O} {Middleware} for
  {Location}-{Flexible} {Scientific} {Data} {Analytics}}, in:
  \bibinfo{booktitle}{2013 {IEEE} 27th {International} {Symposium} on
  {Parallel} and {Distributed} {Processing}}, pp. \bibinfo{pages}{320--331}.
\newblock \DOIprefix\doi{10.1109/IPDPS.2013.46}. \bibinfo{note}{iSSN:
  1530-2075}.

\end{thebibliography}

\begin{appendix}

\section{Interpolation with Co-moving Coordinates}\label{sec:comoving}

The Hermite interpolation between time $t_0$ and $t_1$ at $t$ is given by:

\begin{equation}
    p_n(u) = \sum_{i=0}^{(n-1)/2} p_0^{(i)}H_i^n(u) + p_1^{(i)} H_{n-i}^n(u)
\end{equation}
where $n$ is the degree of the interpolation, $p_0^{(i)}$ ($p_1^{(i)}$) is the
i-th derivative at $t_0$ ($t_1$) and $u$ is the normalized time variable ($(t -
t_0) / (t_1 - t_0)$).  The polynomials $H_i^n$ are given by the line of the
following matrix where the first column corresponds to the constant term

\citep{lind_real-time_2020}:
  \begin{equation}
    H^3 = \left(
    \begin{array}{cccc}
      1 & 0 & -3 & 2 \\
      0 & 1 & -2 & 1 \\
      0 & 0 & -1 & 1 \\
      0 & 0 & 3 & -2
    \end{array} \right).
  \end{equation}
  For example $H_0^3(u)$ is given by $1 - 3 u^2 + 2 u^3$.
  For the quintic interpolation, the matrix is given by:
  \begin{equation}
    H^5 = \left(
    \begin{array}{cccccc}
      1 & 0 & 0 & -10 & 15 & -6 \\
      0 & 1 & 0 & -6 & 8 & -3 \\
      0 & 0 & \frac{1}{2} & -\frac{3}{2} & \frac{3}{2} & -\frac{1}{2} \\
      0 & 0 & 0 & \frac{1}{2} & -1 & \frac{1}{2} \\
      0 & 0 & 0 & -4 & 7 & -3 \\
      0 & 0 & 0 & 10 & -15 & 6
    \end{array}
  \right).
  \end{equation}

In cosmological simulation, co-moving coordinates are required to take into
account the expansion of the universe (for more information on co-moving
coordinates see for example
\cite{bertschinger_simulations_1998,peebles_principles_1993}).  In \swift, they
are defined as $x=ax'$, $a^2\dot{x}' = v'$ and $a \dot{v}' = g'$\footnote{The
last one holds only for gravity as hydrodynamics have a different relation.}
where $x'$, $v'$ and $g'$ are the co-moving coordinates, velocities and
accelerations, and $a$ is the scaling factor that depends on time\footnote{Size
of the universe normalized to today's value.} .  To use the Hermite
interpolation previously described, $u$ is replaced by $(a - a_0) / (a_1 - a_0)$
and the terms $p_0^{(i)}$ and $p_1^{(i)}$ need to include some cosmological
factors.  Those last terms are obtained from the derivation of the co-moving
coordinate with respect to the scale factor.  For the positions, the velocities
$p_0^{(1)}$ and $p_1^{(1)}$ need to be modified with the following expression
evaluated at either $a_0$ or $a_1$:
  \begin{equation}
    \frac{\der x'}{\der a} = \frac{\der x'}{\der t} \frac{\der t}{\der a} = \frac{v'}{a^2 \dot{a}}.
  \end{equation}
  The derivative of $a$ with respect to time is given by:
  \begin{equation}
    \dot{a} = a H_0 \sqrt{(\Omega_c + \Omega_b)a^{-3} + \Omega_{rad}a^{-4} + \Omega_\Lambda}
  \end{equation}
  where $\Omega$ are the cosmological parameters and $H_0$ the Hubble constant.
  This equation assumes a flat $\Lambda$-CDM universe.
  For the accelerations within the position interpolation ($p_0^{(2)}$ and $p_1^{(2)}$),
  an additional derivative is required:
  \begin{equation}
    \frac{\der^2 x'}{\der a^2} = \frac{1}{a^4 \dot{a}^2}\left( ag' - 2a\dot{a}v' - \frac{a^2\ddot{a}}{\dot{a}}v' \right).
  \end{equation}

In the case of the interpolation of velocities, the same computation is required
for the accelerations ($p_0^{(1)}$ and $p_1^{(1)}$) and gives:
\begin{equation}
  \frac{\der v'}{\der a} = \frac{g'}{\dot{a}a}
\end{equation}
 
\section{Example of python scripts}

  \ownsummary{Aquamarine}{
    Show how to use the python wrapper.
  }
  The following code shows a quick example of the \csds's interface.
  The full interface is described in the documentation provided within the
  reader's repository.

  \onecolumn
  \pythonexternal[caption=Example of the \csds's interface.]{example.py}

\end{appendix}

\label{lastpage}
\end{document}